\documentclass[]{aa}  

\usepackage{graphicx}
\usepackage{txfonts}
\usepackage{amssymb}
\usepackage{multirow}
\usepackage{float}
\usepackage{gensymb}
\usepackage{xcolor}
\usepackage{verbatim} 
\usepackage{rotating}
\usepackage{sidecap}
\usepackage[]{hyperref}
\hypersetup{backref=true, pagebackref=true, hyperindex=true, breaklinks=true,colorlinks=true,urlcolor=blue, linkcolor=blue, citecolor=blue,pagecolor=red, bookmarks=true, bookmarksopen=true}

\newcommand{\kms}{\ensuremath{\rm km\,s^{-1}}\xspace}

\begin{document}

   \title{ALMA chemical survey of disk-outflow sources in Taurus (ALMA-DOT)}

   \subtitle{VI: Accretion shocks in the disk of DG Tau and HL Tau}

   \author{A.\,Garufi \inst{\ref{Firenze}}
   \and L.\,Podio \inst{\ref{Firenze}}
   \and C.\,Codella \inst{\ref{Firenze}, \ref{IPAG}} 
   \and D.\,Segura-Cox \inst{\ref{MPIEP}}
   \and M.\,Vander Donckt \inst{\ref{ESO_Germany}}
   \and S.\,Mercimek \inst{\ref{Firenze}, \ref{Uni_Firenze}}
   \and \\ F.\,Bacciotti \inst{\ref{Firenze}}
   \and D.\,Fedele \inst{\ref{Firenze}}
   \and M.\,Kasper \inst{\ref{ESO_Germany}}
   \and J.\,E.\,Pineda \inst{\ref{MPIEP}}
   \and E.\,Humphreys \inst{\ref{ALMA_Santiago}, \ref{ESO_Germany}}
   \and L.\,Testi\inst{\ref{ESO_Germany}, \ref{Firenze}}
   }

  \institute{INAF, Osservatorio Astrofisico di Arcetri, Largo Enrico Fermi 5, I-50125 Firenze, Italy. \label{Firenze}
  \email{antonio.garufi@inaf.it}  
  \and Univ.\ Grenoble Alpes, CNRS, Institut de Plan\'{e}tologie et d'Astrophysique de Grenoble (IPAG), 38000 Grenoble, France \label{IPAG}
  \and Center for Astrochemical Studies, Max Planck Institute for Extraterrestrial Physics, Garching, 85748, Germany \label{MPIEP}
  \and European Southern Observatory, Karl-Schwarzschild-Strasse 2, D-85748 Garching, Germany \label{ESO_Germany}
  \and Universit\`a degli Studi di Firenze, Dipartimento di Fisica e Astronomia, Via G. Sansone 1, 50019 Sesto Fiorentino, Italy \label{Uni_Firenze}
  \and Joint ALMA Observatory, Av. Alonso de Cordova 3107, Vitacura, Santiago, Chile \label{ALMA_Santiago}
  }

   \date{Received -; accepted -}

 
 \abstract{Planet-forming disks are not isolated systems. Their interaction with the surrounding medium affects their mass budget and chemical content. In the context of the ALMA-DOT program, we obtained high-resolution maps of assorted lines from six disks that are still partly embedded in their natal envelope. In this work, we examine the SO and SO$_2$ emission that is detected from four sources: \mbox{DG Tau}, HL Tau, IRAS 04302+2247, and T Tau. The comparison with CO, HCO$^{+}$, and CS maps reveals that the SO and SO$_2$ emission originates at the intersection between extended streamers and the planet-forming disk. Two targets, DG Tau and HL Tau, offer clear cases of inflowing material inducing an accretion shock on the disk material. The measured rotational temperatures and radial velocities are consistent with this view. In contrast to younger Class 0 sources, these shocks are confined to the specific disk region impacted by the streamer. In HL Tau, the known accreting {streamer} induces a shock in the disk outskirts, and the released SO and SO$_2$ molecules spiral toward the star in a few hundred years. These results suggest that shocks induced by late accreting material may be common in the disks of young star-forming regions with possible consequences for the chemical composition and mass content of the disk. They also highlight the importance of SO and SO$_2$ line observations in probing {accretion} shocks from a larger sample.}

\authorrunning{Garufi et al.}

\titlerunning{ALMA-DOT VI. Accretion shocks in the disk of DG Tau and HL Tau}

   \maketitle
%

\section{Introduction}
Planetary systems form in protoplanetary disks around protostars through the assembly of gas and dust particles. Their architecture is inescapably influenced by the disk size and mass at the time of planet formation. Accretion onto the protostar and protoplanetary disk proceeds through the funnelling of material from the collapsing envelope surrounding the protostar \citep{Terebey1984} and, for an isolated system, is expected to last as long as there is sufficient material shrouding the system \citep[$\sim10^5$ yr,][]{Machida2010}. However, the accretion timescale of cloud material onto the protostellar system is severely affected by the large-scale environment \citep{Padoan2014, Kuffmeier2017} and can span up to an order of magnitude for similarly massive stars. Episodic accretion prolonged over time can explain both the observed luminosity spread of star-forming regions \citep{Baraffe2009} and the stellar luminosity bursts imprinted in the disk chemical properties \citep{Jorgensen2015}. An interesting consequence of late accretion events is that the {total} mass budget available for the planet formation exceeds the mass measured in a protoplanetary disk at any given time {because this is replenished, which provides} a possible solution to the missing mass problem \citep{Manara2018, Kuffmeier2020}. 

The (sub-)millimeter high-resolution imaging enabled by the Atacama Large {Millimeter/submillimeter} Array (ALMA) has brought to light both the dusty and gaseous structure of the protostellar and protoplanetary environment. The existence of substructures in the protoplanetary disk  early in their evolution \citep{ALMA2015} suggests that the planet formation might already be underway in the first $10^6$ yr of the life of a star. {In such young stars,} the emission from the envelope is still dominant over that from the protostar, and {their} spectral energy distribution (SED) peaks at {far}-infrared (FIR) wavelengths. These stars are observationally defined as Class I following \citet{Lada1987}, and are precursors of the most studied class, Class II, where the envelope has dissipated but the disk still accretes onto the star. The first, current census of Class I sources reveals the high occurrence of both dust substructures in the disk \citep{Sheehan2017, Sheehan2020, Segura-Cox2020} and complex non-Keplerian gaseous structures around the disk \citep{Fernandez-Lopez2020, Huang2020, Alves2020}. These two elements combined determine our need to characterize the interaction of disks with the environment in this type of object in order to constrain both the available mass budget and the physical processes at play when planets form. Of particular interest in this regard is the recent discovery of extended streamers feeding the central protostar or disk. These streamers are revealed with diverse molecular tracers, such as carbon monoxide \citep[CO,][]{Akiyama2019}, formyl cation \citep[HCO$^+$,][]{Yen2019}, and cyanoacetylene \citep[HC$_3$N,][]{Pineda2020} which emphasizes the importance of datasets of assorted molecular lines.  

The chemical characterization of partly embedded sources is the main goal of the ALMA chemical survey of disk-outflow sources in Taurus \citep[ALMA-DOT,][]{Garufi2021}. In the context of this campaign, we obtained ALMA high-resolution images of 25 spectral lines of nine molecular species toward six Class I or Class I/II sources. The previous papers of this series focus on the disk molecular emission from specific targets \citep{Podio2020, Podio2020b, Garufi2020b} or on specific molecules \citep{Podio2019, Codella2020}. This is the sixth paper of the series and is devoted to the characterization of extended structures around the disks and to the line emission of sulfur monoxide (SO) and sulfur dioxide (SO$_2$ and $^{34}$SO$_2$).

Sulfur-bearing species like SO and SO$_2$ are typically observed in shocked regions {along protostellar jets and outflows \citep[e.g., ][]{Bachiller1997,Lee2010,Tafalla2010,Codella2014b,Podio2015,Podio2021}}. These {shocks} are perfect laboratories with which to study the enrichment of the chemical content of star forming regions thanks to sputtering (gas-grain collisions) and shattering (grain-grain) processes {which cause the release of the grain mantles and cores 
into the gas phase} \citep[e.g.,][and references therein]{Flower1994,Gusdorf2008a,Gusdorf2008b,Guillet2011}. In addition, the sudden increase in temperature and density {in shocks} triggers a hot gas-phase chemistry, which is otherwise not efficient in more quiescent and colder regions.
As a consequence, the abundance of several molecular species in addition to S-bearing ones dramatically increases by several orders of magnitude \citep[e.g.,][]{Bachiller1997,Bachiller2001,Codella2005,Jimenez2005,Jorengensen2007}.
The main S-reservoir on dust mantles is still unknown.
It has been postulated to be H$_2$S, but this has never {been} detected in interstellar ices \citep[e.g.,][]{Charnley1997,Boogert2015,Laas2019}.
 SO and SO$_2$ are {between} the most abundant S-bearing species associated with fast ($\geq$ 10 km s$^{-1}$) shocks driven by the propagation of supersonic protostellar jets \citep{Bachiller1997,Tafalla2010}. 
{In addition to fast shocks along outflows}, slow ($\sim$ 1 km s$^{-1}$) shocks {have} recently been observed {at} the centrifugal barrier, which is the transition region between the infalling rotating envelope and the accretion disk \citep[e.g.,][]{Sakai2014,Sakai2017,Oya2016,Lee2019,Codella2019}. These slow shocks {do not} affect the refractory core {of dust grains} but {sputter their icy mantles, enhancing the gas-phase abundance of several molecules}. Again, sulphur-bearing species play a major role in this context, and indeed the imaging of SO emission towards L1527 using ALMA \citep{Sakai2014} paved the way for studies of the {so-called accretion shocks, which occur} where the {infalling} envelope {impacts onto the disk}.

\begin{table}
 \centering
 \caption{Molecular lines probed.}
 \label{Line_table}
  \begin{tabular}{ccccc}
  \hline
  Molecule & Transition & $\nu_{\rm rest}$ & E$\rm _{up}$ & S$_{\rm ij}\mu^2$ \\
   & & (GHz) & (K) & (D$^2$) \\
   \hline
   \multicolumn{5}{c}{Band 5 -- DG Tau} \\ 
   \multicolumn{5}{c}{Cycle 5, 2017.1.01562.S} \\ 
   \hline
SO$_2$ & $2_{2,0}-1_{1,1}$ & 192.65102 & 13 & 3.9 \\
SO$_2$ & $9_{1,9}-8_{0,8}$ & 193.60949 & 42 & 15.9 \\
SO$_2$ & $12_{0,12}-11_{1,11}$ & 203.39155 & 70 & 22.5 \\
SO$_2$ & $18_{3,15}-18_{2,16}$ & 204.24676 & 181 & 34.6 \\
SO$_2$ & $7_{4,4}-8_{3,5}$ & 204.38430 & 65 & 1.7 \\
    \smallskip
SO     & $4_{5}-3_{4}$ & 206.17601 & 39 & 8.9 \\
    \hline
   \multicolumn{5}{c}{Band 6 -- DG Tau, HL Tau, IRAS04302, T Tau} \\
   \multicolumn{5}{c}{Cycle 4 and 6, 2016.1.00846.S and 2018.1.01037.S} \\
   \hline
SO$_2$ & $11_{5,7}-12_{4,8}$ & 229.34763 & 122 & 3.1 \\
SO$_2$ & $5_{2,4}-4_{1,3}$ & 241.61579 & 24 & 5.7 \\ 
SO$_2$ & $5_{4,2}-6_{3,3}$ & 243.08764 & 53 & 0.7 \\
SO$_2$ & $14_{0,14}-13_{1,13}$ & 244.25421 & 93 & 28.0 \\
    \smallskip
    $^{34}$SO$_2$ & $4_{2,2}-3_{1,3}$ & 229.85761 & 19 & 4.6 \\ 
    \smallskip
CO         & $2-1$ & 230.53800 & 17 & 0.02 \\
    \smallskip
o-H$_2$CO  & $3_{1,2}-2_{1,1}$ & 225.69777 & 33 & 43.5 \\
    \smallskip
CS & $5-4$ & 244.93555 & 35 & 19.1 \\
    \smallskip
HCO$^+$ & $3-2$ & 267.55763 & 26 & 45.6 \\
    \hline
   \multicolumn{5}{c}{Band 7 -- HL Tau} \\ 
   \multicolumn{5}{c}{Cycle 5, 2017.1.01562.S} \\ 
   \hline
    \smallskip
SO$_2$ & $3_{3,1}-2_{2,0}$ & 313.27972 & 28 & 6.7 \\
   \hline
   \multicolumn{5}{c}{Band 9 -- HL Tau} \\ 
   \multicolumn{5}{c}{Cycle 5, 2017.1.01562.S} \\ 
   \hline
SO     & $14_{15}-13_{14}$ & 644.37892 & 254 & 32.9 \\
SO     & $15_{15}-14_{14}$ & 645.25493 & 261 & 35.2 \\
    \smallskip
SO     & $16_{15}-15_{14}$ & 645.87592 & 253 & 37.6 \\
  \hline
   \end{tabular}
   \tablefoot{Columns are: molecular species, transition, frequency at rest frame, upper-level energy, and line strength. All parameters are from CDMS \citep{Mueller2005}. {The HCO$^+$ data are from \citet{Yen2019}.}}
\end{table}

In this work, we report the detection of SO and SO$_2$ from the class I/II DG Tau {\citep[at 125.3 pc,][]{Gaia2021}} and HL Tau {\citep[at 147.3 pc,][]{Galli2018}} and of SO$_2$ from the Class I sources IRAS 04302+2247 {\citep[hereafter IRAS04302, at 161 pc,][]{Galli2019}} and T Tau {\citep[at 145.1, pc][]{Gaia2021}}. The SO and SO$_2$ emission is spatially connected with faint streamers visible in various gaseous tracers. The paper is organized as follows. In Sect.\,\ref{Observations}, we describe the observing setup and the data reduction. In Sect.\,\ref{Results} we present the results of the analysis, and in Sects.\,\ref{Discussion} and \ref{Conclusions} we discuss our findings and present our conclusions.

\begin{table*}
 \centering
 \caption{Observed lines properties and integrated fluxes.}
 \label{Line_setup}
  \begin{tabular}{ccccccccc}
  \hline
  Molecule & Transition & $\Delta$V & Briggs & r.m.s.  & Beam size & Flux$_{\rm disk}$ & Flux$_{\rm beam}$ & Flux$_{\rm beam}$ \\
           &            & (km s$^{-1}$) &    & (mJy beam$^{-1}$) & (\arcsec) & (mJy km s$^{-1}$) & (mJy km s$^{-1}$) & (mJy km s$^{-1}$)\\
  \hline
\multicolumn{7}{c}{DG Tau -- Band 5} & Outer disk &  \\
SO$_2$ & $2_{2,0}-1_{1,1}$        & 0.8  & 2.0 & 0.8 & 0.67$\times$0.49 & $<18$ & $<6$    &\\
SO$_2$ & $9_{1,9}-8_{0,8}$        & 0.8  & 2.0 & 0.6 & 0.66$\times$0.48 & $<16$ & $<6$    &\\
SO$_2$ (*) & $12_{0,12}-11_{1,11}$& 0.8  & 2.0 & 0.6 & 0.64$\times$0.47 & (41)  & $8\pm6$ &\\
SO$_2$ & $18_{3,15}-18_{2,16}$    & 0.8  & 2.0 & 0.7 & 0.68$\times$0.49 & $<20$ & $<6$    &\\
SO$_2$ & $7_{4,4}-8_{3,5}$    & 0.8  & 2.0 & 0.8 & 0.64$\times$0.46 & $<18$ & $<6$    &\\ 
\smallskip
\smallskip
SO (*)     & $4_{5}-3_{4}$            & 0.8  & 0.0 & 0.7 & 0.41$\times$0.29 & 285   & $75\pm6$&\\
\multicolumn{7}{c}{DG Tau -- Band 6} & &  \\
SO$_2$ & $11_{5,7}-12_{4,8}$      & 0.16 & 0.0 & 1.7 & 0.14$\times$0.11 & $<28$ & $<6$ &\\
SO$_2$ & $5_{2,4}-4_{1,3}$        & 0.16 & 0.0 & 1.7 & 0.13$\times$0.11 & $<33$ & $<6$ &\\ 
SO$_2$ & $5_{4,2}-6_{3,3}$        & 0.16 & 0.0 & 1.8 & 0.13$\times$0.11 & $<25$ & $<6$ &\\
SO$_2$ & $14_{0,14}-13_{1,13}$    & 0.6  & 0.0 & 0.6 & 0.16$\times$0.13 & (26)  & $<6$ &\\
$^{34}$SO$_2$ & $4_{2,2}-3_{1,3}$ & 0.16 & 0.0 & 1.5 & 0.13$\times$0.11 & $<26$ & $<6$ &\\
  \hline
\multicolumn{7}{c}{HL Tau -- Band 6} & Outer disk & Inner disk \\ 
SO$_2$ & $11_{5,7}-12_{4,8}$      & 0.2  & 2.0 & 2.1 & 0.40$\times$0.34 & $<29$ & $<6$    & $<6$\\
SO$_2$ (*) & $5_{2,4}-4_{1,3}$    & 0.2  & 0.0 & 2.0 & 0.30$\times$0.25 & 282   & $30\pm6$& $<6$\\
SO$_2$ & $5_{4,2}-6_{3,3}$        & 0.2  & 2.0 & 2.2 & 0.37$\times$0.32 & $<19$ & $<6$    & $<6$\\
SO$_2$ (*) & $14_{0,14}-13_{1,13}$& 1.2  & 0.0 & 0.9 & 0.29$\times$0.27 & 441   & $46\pm9$& $25\pm9$\\
\smallskip
\smallskip
$^{34}$SO$_2$ & $4_{2,2}-3_{1,3}$ & 0.2  & 2.0 & 2.2 & 0.40$\times$0.34 & $<18$ & $<6$    & $<6$\\
\multicolumn{7}{c}{HL Tau -- Band 7} & & \\ 
\smallskip
\smallskip
SO$_2$ & $3_{3,1}-2_{2,0}$        & 1.1  & 2.0 & 1.0 & 0.13$\times$0.09 & 467   & $21\pm5$ & $<5$ \\
\multicolumn{7}{c}{HL Tau -- Band 9} &  & \\ 
SO     & $14_{15}-13_{14}$        & 0.25 & 2.0 & 9.4 & 0.14$\times$0.10 & 8722  & $151\pm90$ & $543\pm90$\\
SO     & $15_{15}-14_{14}$        & 0.25 & 2.0 & 12.4& 0.15$\times$0.10 & 7010  & $<90$      & $591\pm90$\\
SO (*) & $16_{15}-15_{14}$        & 0.25 & 2.0 & 13.8& 0.15$\times$0.10 & 10351 & $242\pm90$ & $653\pm90$\\
  \hline
\multicolumn{7}{c}{IRAS04302 -- Band 6} & Outer disk & Inner disk  \\
SO$_2$ & $11_{5,7}-12_{4,8}$      & 0.2 & 2.0 & 2.2 & 0.43$\times$0.34 & $<22$  & $<6$    & $<6$\\
SO$_2$ & $5_{2,4}-4_{1,3}$        & 0.2 & 2.0 & 2.2 & 0.41$\times$0.32 & (52)   & $7\pm6$ & $<6$\\
SO$_2$ & $5_{4,2}-6_{3,3}$        & 0.2 & 2.0 & 2.0 & 0.40$\times$0.32 & $<25$  & $<6$    & $<6$\\
SO$_2$ (*) & $14_{0,14}-13_{1,13}$& 1.2 & 2.0 & 0.9 & 0.41$\times$0.34 & (47)   & $<6$    & $12\pm6$\\
$^{34}$SO$_2$ & $4_{2,2}-3_{1,3}$ & 0.2 & 2.0 & 2.3 & 0.43$\times$0.33 & $<24$  & $<6$    & $<6$\\
  \hline
\multicolumn{7}{c}{T Tau S -- Band 6} & Disk & Knot \\
SO$_2$ & $11_{5,7}-12_{4,8}$      & 0.2 & 2.0 & 1.9 & 0.41$\times$0.34 & $<25$  & $<9$    & $<9$     \\
SO$_2$ (*) & $5_{2,4}-4_{1,3}$        & 0.2 & 0.0 & 2.2 & 0.30$\times$0.24 & 411    & $16\pm9$& $17\pm9$ \\
SO$_2$ & $5_{4,2}-6_{3,3}$        & 0.2 & 2.0 & 2.2 & 0.38$\times$0.32 & $<27$  & $<9$    & $<9$     \\
SO$_2$ (*) & $14_{0,14}-13_{1,13}$& 1.2 & 0.0 & 0.9 & 0.29$\times$0.26 & 416    & $79\pm9$& $16\pm9$ \\
$^{34}$SO$_2$ & $4_{2,2}-3_{1,3}$ & 0.2 & 2.0 & 2.2 & 0.38$\times$0.32 & $<22$  & $<9$    & $<9$     \\
   \hline
   \end{tabular}
   \tablefoot{Columns are: molecular species, transition, channel width, Briggs weighting robustness, r.m.s., beam size, {disk-integrated flux, and beam-integrated fluxes over the regions where the gas temperature and molecular column density were derived}. The detected lines used to infer $N$ and $T_{\rm rot}$ are marked with an asterisk. {Brackets denote tentative detections, i.e., disk-integrated intensities between 3$\sigma$ and 5$\sigma$. For nondetected lines, we report upper limits equivalent to $3\sigma$.}}
\end{table*}

\section{Observations and data reduction} \label{Observations}
This work makes use of ALMA Band 5, 6, 7, and 9 observations from Cycles 4, 5, and 6. Within the ALMA-DOT program (P.I.: Podio, L.), {six sources were observed in Band 6 during Cycle 4 (2016.1.00846.S) and Cycle 6 (2018.1.01037.S) using 12 narrow and high-resolution ($0.16$ km\,s$^{-1}$ and $0.2$ km\,s$^{-1}$)  spectral windows (SPWs) and one broad and low-resolution SPW ($0.6$ km\,s$^{-1}$ and $1.2$ km\,s$^{-1}$) for the continuum. An overview of the ALMA-DOT dataset was presented by \citet{Garufi2021}, while in the context of this paper we focus on the} five spectral windows (SPWs) covering four SO$_2$ transitions and one $^{34}$SO$_2$ transition, {and on the four sources that show emission in SO$_2$, namely DG Tau, HL Tau, IRAS04302, and T Tau}. {In addition to the Band 6 observations,} DG Tau was observed in Band 5 during Cycle 5 (2017.1.01562.S) {using 12 narrow and high-resolution SPWS ($0.8$  km\,s$^{-1}$), covering four SO$_2$ transitions and one SO transition, and one broad and low-resolution SPW for the continuum.} 
We also make use of ALMA {Cycle 5} observations {of HL Tau} from program 2017.1.01178.S (P.I.: Humphreys, E.) covering one SO$_2$ transition in Band 7 and three SO transitions in Band 9. In total, 11 SO$_2$ lines (including the $^{34}$SO$_2$ isotopolog) and 4 SO lines are probed by the observations presented in this work. The properties of these lines, based on the parameters from the Cologne Database for Molecular Spectroscopy \citep[CDMS,][]{Mueller2005} are listed in Table
\ref{Line_table}.

Data reduction was performed with the Common Astronomy Software Applications package \citep[CASA,][]{McMullin2007} version 4.7.2 for the ALMA-DOT programs, and 5.7.2 for 2017.1.01178.S. Self-calibration was applied to the strong continuum emission improving the S/N of the continuum image by a factor of 2.4, 3.3, 3.4, and 4.5 for DG Tau, HL Tau, IRAS04302, and T Tau, respectively. These solutions were applied to the line-free continuum-subtracted SPWs. Spectral cubes were produced using \textsc{tclean} interactively through a manually selected mask on the visible signal. Maps were generated with a Briggs robustness of both 0.0 and 2.0. A posteriori, we adopted a value of 0.0 for the strong detections in order to maximize the angular resolution, and of 2.0 for the faint and nondetections to maximize the recovered flux, and for the 2017.1.01178.S high-angular-resolution data. {The beam sizes of the final images span from 0.12\arcsec\ to 0.58\arcsec\ while the root mean square (r.m.s.) noise per channel goes from 0.6 to 13.8 mJy beam$^{-1}$}. The {properties of the obtained line cubes (spectral resolution, Briggs robustness, r.m.s per channel, and clean beam) of the obtained datacubes} of all SO$_2$ and SO line observations are shown {in Table \ref{Line_setup}}.

\begin{figure*}
  \centering
 \includegraphics[width=18cm]{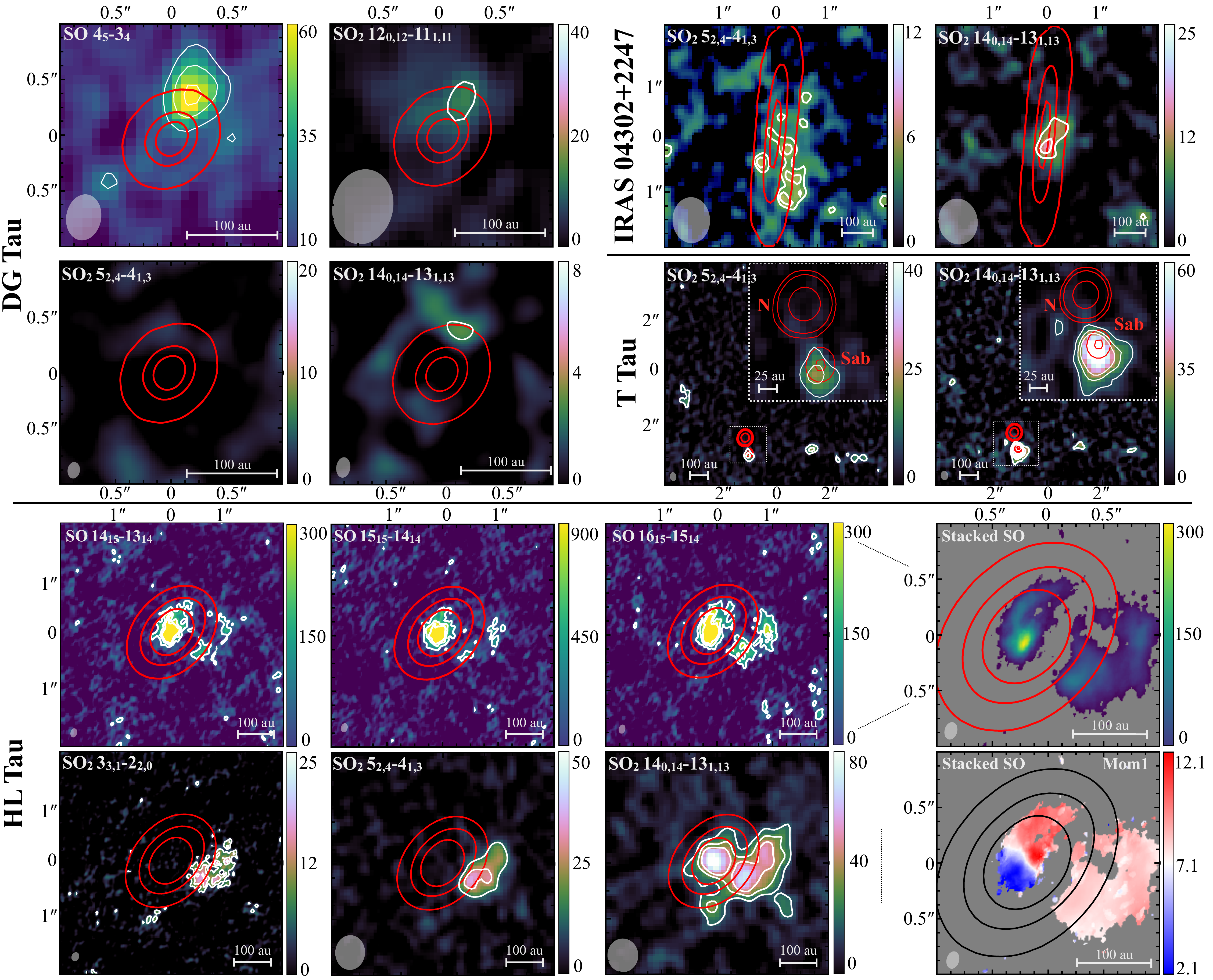} 
     \caption{Moment-0 maps of SO and SO$_2$ emission. The red and white contours indicate the continuum {at 1.3~mm} and line emission, respectively. The continuum contours of DG Tau are at 15$\sigma$, 90$\sigma$, and 160$\sigma$ significance while the line contours at 4$\sigma$, 8$\sigma$, and 12$\sigma$; the continuum contours of IRAS04302 are at 14$\sigma$, 200$\sigma$, and 400$\sigma$ while the line contours are at 3$\sigma$ and 4$\sigma$; the continuum contours of T Tau are at 50$\sigma$, 140$\sigma$, and 1000$\sigma$ while the line contours are at 3$\sigma$, 5$\sigma$, and 8$\sigma$; the continuum contours of HL Tau are at 8$\sigma$, 40$\sigma$, and 80$\sigma$ while the SO$_2$ line contours are at 5$\sigma$, 10$\sigma$, and 15$\sigma$ and SO line contours are at 3$\sigma$, 6$\sigma$, and 9$\sigma$. A zoom onto T Tau N and S is shown in the inset image of the source. Panels in the last column of HL Tau are the zoomed-in moment-0 and moment-1 maps of the stacked SO emission. {We note that only contours at $\geq5\sigma$ denote formal detection.} The beam size is indicated by the gray ellipse to the bottom left. Color units are mJy beam$^{-1}$ km s$^{-1}$ in the moment-0 and km s$^{-1}$ in the moment-1 maps, respectively. North is up, east is left.} 
 \label{Imagery_SO2}
 \end{figure*}

Cycle 4 and 6 observations also include molecular lines of $^{12}$CO, {H$_2$CO}, and CS. These two datasets were presented in detail by \citet{Podio2019, Podio2020,Podio2020b} and \citet{Garufi2020b, Garufi2021}, who focused on the molecular emission from the disk. In the present work, we make use of the same dataset to investigate the ambient emission around the four targets with observable SO$_2$ and SO emission (DG Tau, HL Tau, IRAS04302, and T Tau). Readers can find details on the $^{12}$CO, {H$_2$CO}, and CS line properties {in Table \ref{Line_table} and on their} observing settings in the referenced papers. {Finally, we make use of the HCO$^+$ $3-2$ line observations of HL Tau reduced and analyzed by \citet{Yen2019}. The line properties can be found in Table \ref{Line_setup} and the technical setup in \citet{Yen2019}}.

\section{Data analysis and results} \label{Results}

\subsection{Detected SO and SO$_2$ lines} \label{Morphology}
{To survey the SO and SO$_2$ emission in the disks observed by the ALMA-DOT program we inspected the line cubes of the targeted SO and SO$_2$ lines (see Table \ref{Line_table}). For the four disks where emission is detected in at least one of the SO and SO$_2$ lines, namely DG Tau, HL Tau, IRAS04302, and T Tau, we produce velocity-integrated intensity  (moment-0) maps of the lines, integrating over the channels where emission is detected at $>3\sigma$ (the r.m.s. noise per channel of the line cubes is given in Table \ref{Line_setup}). For the lines showing no emission at the line cube inspection, we integrate over the same velocity range of the detected line(s). Finally, we integrate the moment-0 maps over the disk area (as defined by the area where the 1.3~mm flux is $>5\sigma$) to obtain the disk-integrated line flux summarized in Table \ref{Line_setup}.
Based on the disk-integrated line intensities, three} of the {ten} SO$_2$ lines probed are detected {above 5$\sigma$ confidence} in at least one source, {one is tentatively detected ($3-5\sigma$)}, {while the remaining six, as well as the $^{34}$SO$_2$ line, are never detected ($<3\sigma$)}. The {four SO$_2$ lines formally or tentatively detected} are the $12_{0,12}-11_{1,11}$ (only surveyed and tentatively detected in DG Tau), the $3_{3,1}-2_{2,0}$ (only surveyed in HL Tau), the $5_{2,4}-4_{1,3}$ (detected in HL Tau, IRAS04302, and T Tau), and the $14_{0,14}-13_{1,13}$ (detected in all sources, although only tentatively in DG Tau and IRAS04302). The SO $4_{5}-3_{4}$ line, which is only probed in DG Tau, as well as the {SO} $14_{15}-13_{14}$, $15_{15}-14_{14}$, and $16_{15}-15_{14}$ lines, which are only probed in HL Tau, are all firmly detected. 
Figure \ref{Imagery_SO2} shows the moment-0 maps of the detected SO and SO$_2$ lines {compared with the distribution of the disk continuum emission at 1.3~mm}.

\subsubsection{Spatial distribution of SO and SO$_2$ emission} \label{Emission_morphology}
Figure \ref{Imagery_SO2} {shows that} the strong SO $4_5-3_4$ line emission in DG Tau  {is not located along the direction of its well-known collimated jet, which is detected along $PA_{\rm jet} = 225\degree$, i.e., perpendicular to the position angle (PA) of the disk ($PA_{\rm disk} = 135\degree$) \citep[e.g., ][]{Eisloffel1998,Podio2020b}. Moreover, the SO emission is not symmetrically distributed across the disk but instead originates  from only one side of the disk, the northwest side along the disk major axis. The peak of the emission, detected} at 12$\sigma$, {is located} at 0.4\arcsec\ (50 au) {to the northwest with respect to} the continuum peak. The SO$_2$ $12_{0,12}-11_{1,11}$ line {is tentatively detected (3$\sigma$) and peaks at the same position {as SO}. The SO$_2$ $14_{0,14}-13_{1,13}$ line is not detected although some marginal flux is visible at the same location after smoothing the image by 1\arcsec}. 

In HL Tau, the SO$_2$ 5$_{2,4}-4_{1,3}$, $14_{0,14}-13_{1,13}$, {and $3_{3,1}-2_{2,0}$} lines are promptly detected with fluxes up to 13$\sigma$, 17$\sigma$, {and 18$\sigma$}, respectively. Similarly to what is found for DG Tau, {the SO$_2$ emission is not located along the collimated atomic jet direction ($PA_{\rm jet} = 51\degree$, e.g., \citealt{Mundt1990}), or along the wide-angle outflow cavities probed by CO $1-0$ \citep{ALMA2015}.}
The {low-excitation} {SO$_2$ 5$_{2,4}-4_{1,3}$ and $3_{3,1}-2_{2,0}$ line emission ($E_{\rm up} \sim 24$ K and $28$ K, respectively) originates from a compact region on the southwest side of the disk. The emission is displaced to the west with respect to the jet direction and} is displaced from the center of the continuum emission, {extending from a radius of} 0.6\arcsec\ to 1.0\arcsec, {corresponding to radii of} 90 to 150 au. On the other hand, the {high-excitation} {SO$_2$ $14_{0,14}-13_{1,13}$} line {($E_{\rm up} \sim 93$ K)} exhibits both the same displaced component {in the SW outer disk} and a component {from the inner disk, centered on the continuum peak.} 
Similarly to the SO$_2$ $14_{0,14}-13_{1,13}$ line, the three {high-excitation} SO lines {at $\sim$645 GHz ($E_{\rm up} \sim 253-261$ K)} show {bright emission in the inner disk and faint emission from the SW outer disk region}. The moderate-resolution maps by \citet{Wu2018} resolved SO $5_6-4_5$ emission to the SW of {HL Tau}, in line with the SO$_2$ 5$_{2,4}-4_{1,3}$ line emission. Our maps {show} that the central  emission {detected only in the high-excitation SO and SO$_2$ lines}  protrudes toward the north and seems to reconnect, although discontinuously, with the line emission in the {SW outer disk region}. Given their similar E$_{\rm up}$, the three SO lines could be stacked to obtain the maps shown in the last {column of the bottom part of Fig.\,\ref{Imagery_SO2}}. The high spectral and spatial resolution of these data also enables a meaningful intensity-weighted mean velocity (moment-1) map. This map shows that, while the outer component is {detected at a constant redshifted velocity, the inner component shows a velocity gradient that is coherent with the disk rotation pattern} (see Sect.\,\ref{HLTau}).

Only weak SO$_2$ emission is visible from IRAS04302 (Fig.\,\ref{Imagery_SO2}). The {SO$_2$} 5$_{2,4}-4_{1,3}$ line is tentatively detected with 3.5$\sigma$ confidence {in the outer disk} toward the SW of the star and peaks at a separation of 1.2\arcsec\ (190 au). The {SO$_2$} $14_{0,14}-13_{1,13}$ line is {formally} detected (peaking at 5.5$\sigma$) but is only seen close to the continuum center {(i.e. in the inner disk)}.  

\begin{table}
 \centering
 \caption{Beam-averaged column density of SO and SO$_2$ and gas temperature.}
 \label{Line_properties}
  \begin{tabular}{llcc}
  \hline
   Species & Location & $N_{\rm X}$ & $T_{\rm rot}$ \\
   & & (cm$^{-2}$) & (K) \\
  \hline
   SO     & DG Tau outer disk  & $(1.6-6.3) \cdot 10^{14}$  & $40-300$$^a$  \\
   \smallskip
   SO$_2$ & DG Tau outer disk  & $(0.5-2.3) \cdot 10^{14}$  & $40-300$$^a$  \\
   SO    &  HL Tau inner disk  & $>2 \cdot 10^{15}$         & $>350$$^b$    \\
   SO$_2$ & HL Tau inner disk  & $>2 \cdot 10^{15}$         & $>350$$^c$    \\
   SO & HL Tau outer disk      & $(0.2-2) \cdot 10^{16}$    & $58\pm19$$^b$ \\
   \smallskip
   SO$_2$ & HL Tau outer disk  & $(7\pm3) \cdot 10^{14}$    & $58\pm19$     \\
   \smallskip
   SO$_2$ & IRAS04302 inner disk& $>10^{14}$                & $>75$$^c$     \\
   SO$_2$ & T Tau S            & $(0.3-3.7) \cdot 10^{15}$  & $40-300$$^a$  \\
   SO$_2$ & T Tau knot         & $(3\pm2) \cdot 10^{14}$    & 42$\pm$20  \\
   \hline
   \end{tabular}
   \tablefoot{Columns are: molecular species, target and region where the molecular properties are measured, beam-averaged column density, and gas temperature. $^a$: the temperature is assumed. {$^b$: the gas temperature for SO is assumed to be the same as for SO$_2$ as the line emission is cospatial.} $^c$: the lower limit is dictated by the detection {of the SO$_2$ line with} $E_{\rm up}$ = 93 K and the nondetection {of the SO$_2$ line with} $E_{\rm up} = 24$ K.}
\end{table}

Finally, {significant} emission is detected from both the SO$_2$ 5$_{2,4}-4_{1,3}$ and $14_{0,14}-13_{1,13}$ lines {in proximity to} the binary T Tau S (with a flux peak of 8$\sigma$ and 22$\sigma$, respectively). The emission from both lines is displaced from the continuum peak toward the south. {Five to six times} fainter line emission {in the SO$_2$ 5$_{2,4}-4_{1,3}$ and $14_{0,14}-13_{1,13}$ lines} is also detected in a series of knots {located} west of T Tau S {at a distance of $2.5-5\arcsec$, i.e., $\sim 360-720$ au.} No SO$_2$ {emission} is detected {towards} T Tau N.

\subsubsection{Column densities and excitation temperatures} \label{Temperatures}

{To estimate the molecular column density and gas temperature in the SO$_2$- and SO-emitting regions, we integrate the line emission on a beam area centered on the peak of the SO$_2$ $14_{0,14}-13_{1,13}$ line which is detected in all the sources studied here. For HL Tau and IRAS04302, the SO$_2$ emission has two peaks, one in the outer disk and one in the inner disk, and so two integration areas are considered. The same is done for T Tau, where the emission is integrated on the disk of T Tau S and on the closer knot. The beam-integrated line fluxes are summarized in Table \ref{Line_setup}.
For the sources where two or three} lines of the same species {are} detected, the beam-averaged SO and SO$_2$ column density, $N$, and rotational temperature, $T_{\rm rot}$, can be derived from {the beam-integrated line fluxes by performing} a standard rotational diagram {(RD)} assuming local thermodynamic equilibrium (LTE), and optically thin lines.
{The assumption of LTE is well justified as the gas density in the disk molecular layer ($10^{8}-10^{12}$ cm$^{-3}$; e.g., \citealt{Walsh2014}) is above the critical density of the detected SO$_2$ lines ($\sim10^6$ cm$^{-3}$ for kinetic temperatures of $50-350$ K)
and SO lines ($\sim 6 \times 10^5$ cm$^{-3}$ for the $4_5-3_4$ line detected in Band 5 for DG Tau, and $\sim 1.5 \times 10^7$ cm$^{-3}$ for the high $E_{\rm up}$ lines detected in Band 9 for HL Tau, for kinetic temperatures of $50-350$ K)\footnote{the critical densities are inferred using collisional coefficients from the LAMBDA database \citep{Schoier2005}}. 
The line optical depth cannot be estimated as the emission from the $^{34}$SO$_2$ isotopolog is undetected and only up to three lines of SO$_2$ and SO per source are detected, impeding a full radiative transfer analysis. The column densities estimated here assuming optically thin emission have to be considered as lower limits in the case where the lines are optically thick.
For the sources where only one SO$_2$ line is detected (the SO$_2$  $14_{0,14}-13_{1,13}$ line, which has large line strength; see Table \ref{Line_table}), we use the upper limit on the SO$_2$ $5_{2,4}-4_{1,3}$ (which is the line with the second highest line strength in Band 6) to perform a RD analysis and retrieve a lower limit on $T_{\rm rot}$ and $N$. 
The upper limit retrieved for the other undetected SO$_2$ lines in Band 6, which have lower line strengths, is less constraining and is not used in the linear fit of the RD. However, we check that these upper limits are consistent with the obtained solution as shown in the RD plots in Figure \ref{RD}.  
Finally, for DG Tau, where two SO$_2$ lines are both tentatively detected (at $\sim 3\sigma$), and T Tau S, for which the two detected SO$_2$ transitions ($14_{0,14}-13_{1,13}$ and $5_{2,4}-4_{1,3}$) peak at different positions, we assumed a range of temperatures ($T = 40-300$ K) and, under the assumption of LTE and optically thin emission, we estimate a range of column densities from the brightest detected line.
The values of $T_{\rm rot}$ and $N$ obtained for all the sources are reported in Table \ref{Line_properties}.
}

Stringent constraints {on the gas temperature and molecular column density are obtained} for HL Tau, {where the two transitions in Band 6 with the largest line strengths, namely the SO$_2$ 5$_{2,4}-4_{1,3}$ and $14_{0,14}-13_{1,13}$ lines,} are detected {with high S/N}. The rotational temperature {inferred for} the central (inner disk) {region probed only by the high-excitation SO$_2$ line ($E_{\rm up} =93$ K) clearly differs from that inferred for the} displaced (outer disk) component {where also low-excitation SO$_2$ lines are detected ($E_{\rm up} =24-28$ K)} (see Sect.\,\ref{Emission_morphology}). In the {inner disk}, we find a lower limit of 350 K (where the limit is dictated by the nondetection of the 5$_{2,4}-4_{1,3}$ line in the central region) while in the {outer disk} we constrain $T_{\rm rot}=58\pm19$ K. {The estimated SO$_2$ column density is $(7\pm3)\cdot 10^{14}$ cm$^{-2}$ in the outer disk and a factor of two larger in the inner disk.} On the other hand, the {estimate} of the gas temperature {from} the SO {lines} is challenged by the narrow range of $E_{\rm up}$ of the three {detected} lines (253$-$262 K, see Table \ref{Line_table}). {As the SO emission is co-spatial with the SO$_2$ emission both in the outer and inner disk, we assume that the gas temperature of the SO-emitting region is the same as for SO$_2$ (i.e., $58\pm19$ K in the outer disk and $>350$ K in the inner disk) and estimate the SO column density. We find $N_{\rm SO} = (0.2-2) \cdot 10^{16}$ cm$^{-2}$ and $N_{\rm SO}> 2 \cdot 10^{15}$ cm$^{-2}$ for the outer and inner disk region, respectively.}

As for DG Tau, the $T_{\rm rot}$ of the emitting region cannot be constrained from SO or SO$_2$ because of the availability of one SO line and of the faintness of all SO$_2$ lines. {Therefore,} we assumed a {range of temperatures ($T=40-300$ K)} deriving {$N_{\rm SO_2} = (0.5-2.3)\cdot 10^{14}$ cm$^{-2}$, and $N_{\rm SO} = (1.6-6.3)\cdot 10^{14}$ cm$^{-2}$.}

In IRAS04302, we examined the emission from the {inner disk region, where the peak of the brightest SO$_2$ $14_{0,14}-13_{1,13}$} is found. The nondetection of the SO$_2$ 5$_{2,4}-4_{1,3}$ emission towards this region leads to a lower limit on $T_{\rm rot}$ and $N_{\rm SO_2}$ (75 K and 10$^{14}$ cm$^{-2}$, respectively).

Also, {given} the {large uncertainty affecting the} SO$_2$ 5$_{2,4}-4_{1,3}$ emission toward T Tau S {when integrating over the area centered on the brightest SO$_2$ $14_{0,14}-13_{1,13}$ line (see the beam-integrated fluxes reported in Table \ref{Line_setup})} {we assumed a range of temperatures ($T=40-300$ K), as done for DG Tau, and estimate the column density for this temperature range: $N_{\rm SO_2} = (0.3-3.7)\cdot 10^{15}$ cm$^{-2}$}. On the other hand, {two SO$_2$ lines were detected} toward the {brightest} knot located {at $2.5\arcsec$} on the west of T Tau S, {and we were able to constrain}  $T_{\rm rot}$   as 42 K, yielding 
$N_{\rm SO_2}=(3\pm2) \cdot 10^{14}$ cm$^{-2}$.

\subsection{Circumstellar medium from CO, CS, and HCO$^{+}$ lines}
In this section, we examine the ambient medium around the disk of each source  and relate it to the SO and SO$_2$ emission studied in Sect.\,\ref{Morphology}.

\begin{figure*}
  \centering
 \includegraphics[width=18cm]{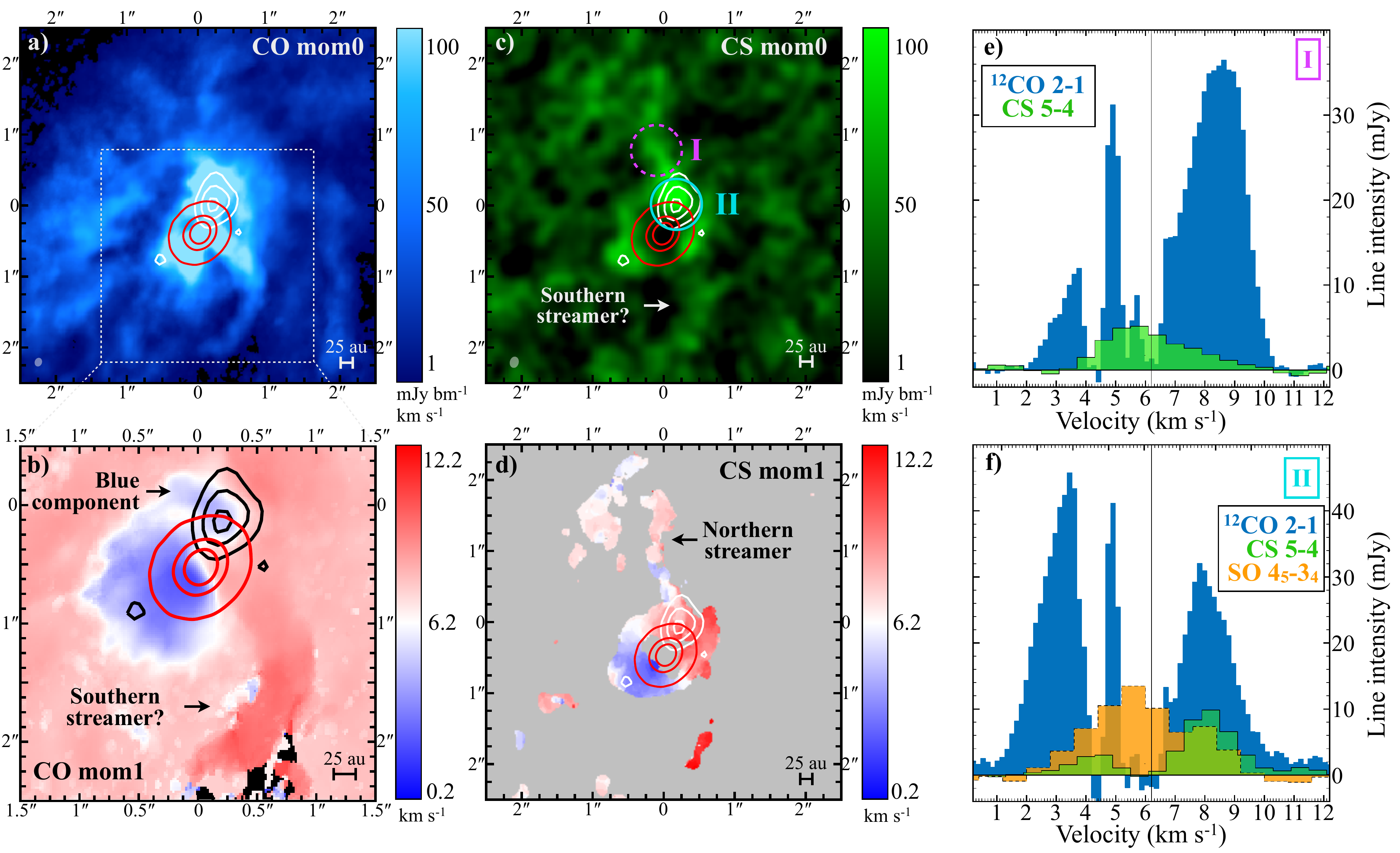} 
     \caption{Imagery of DG Tau. (a): moment-0 map of the $^{12}$CO 2$-$1 line. (b): moment-1 map of the $^{12}$CO 2$-$1 line. (c): moment-0 of the CS 5$-$4 line. (d): moment-1 of the CS 5$-$4 line. (e): Integrated spectra of CO and CS from the region labelled I in (c). (f): Integrated spectra of CO, SO, and CS from the region labelled II in (c). In each map, red and white (or black) contours indicate the continuum and {SO $4_5-3_4$} line emission, respectively, as in Fig.\,\ref{Imagery_SO2}. North is up, east is left.} 
 \label{Imagery_DGTau}
 \end{figure*}

\subsubsection{DG Tau} \label{DGTau}
The $^{12}$CO map of DG Tau (see Fig.\,\ref{Imagery_DGTau}a) reveals the presence of extended structures on a complex kinematics around the disk. A very bright blueshifted component is present on the redshifted (NW) side of the disk. This component is even brighter than the disk itself {(twice as bright as the CO emission on the other side of the disk  at the same separation)}, yielding a blue region in the moment-1 map (Fig.\,\ref{Imagery_DGTau}b). Interestingly, this blue CO component is spatially coincident with the SO flux described in Sect.\,\ref{Morphology}. Another peculiar structure of the CO map is a redshifted structure {resembling a streamer} to the South. Both the blue component and the {putative southern streamer} were detected by \citet{Guedel2018} who explained them with an outflow driven by magnetic fields or a stellar wind.

The CS and H$_2$CO emission from the disk of DG Tau {originates from} a narrow {disk} ring \citep[see][]{Podio2019, Podio2020b}. The {emission} from both molecules is maximized {on the NW side of the disk, which is the region where we detect} the CO blue component and SO {emission}. In Fig.\,\ref{Imagery_DGTau}c, we show a smoothed version of the CS moment-0 map by \citet{Podio2020b}. Two tenuous streamers are visible from both this map \citep[{northern and southern streamers in Fig.\,\ref{Imagery_DGTau}c and d}, see also][]{Podio2020b} and the moment-1 map (Fig.\,\ref{Imagery_DGTau}d) obtained after clipping fluxes below 5$\sigma$.  {The faint southern streamer is spatially coincident with the CO counterpart.} 

{The CS data was analyzed by applying an analytic streamline solution of a rotating sphere collapsing towards a central mass \citep{Mendoza2009,Pineda2020} to search for signs of infall (see Appendix \ref{App_streamers} for a description of the modeling process and Appendix \ref{App_DGTau} for the modeling results of the DG Tau streamers). The results indicate that the northern streamer is indeed infalling and impacting the disk where the SO emission appears, but the analytic streamline model did not well-describe the southern red-shifted arc-shaped streamer. This may be because the southern streamer is actually an extension of the northern streamer piercing through the disk and in the process of turning back towards the central mass.}

Figure\,\ref{Imagery_DGTau}e  shows the spectra of CO and CS {integrated on the brightest part of the} northern streamer {(region I in Fig.\,\ref{Imagery_DGTau}c)}. The CS emission is detected over a relatively large velocity range around the systemic velocity  ($V_{\rm sys} = 6.2$ km s $^{-1}$, \citealt{Garufi2021}) 
but peaks at slightly blueshifted velocities ($\sim$5.5 km s$^{-1}$). From the comparison with CO, it is clear that the northern streamer cannot be promptly detected in CO because of the absorption by the large-scale cloud from 5.2 to 6.6 km s$^{-1}$ (as well as from 3.8 to 4.6 km s$^{-1}$). Finally, Fig.\,\ref{Imagery_DGTau}f reveals that the SO  {spectrum integrated on the NW outer disk component (region II in Fig.\,\ref{Imagery_DGTau}c}) is spectrally broad ($\sim$7 km s$^{-1}$) and peaks at a slightly blueshifted velocity of 5.5 km s$^{-1}$, and is therefore closer in kinematics to the northern streamer {seen in CS} and the CO blue component than to the local redshifted disk (at $\sim$8.5 km s$^{-1}$ from the CS).

\begin{figure*}
  \centering
 \includegraphics[width=18cm]{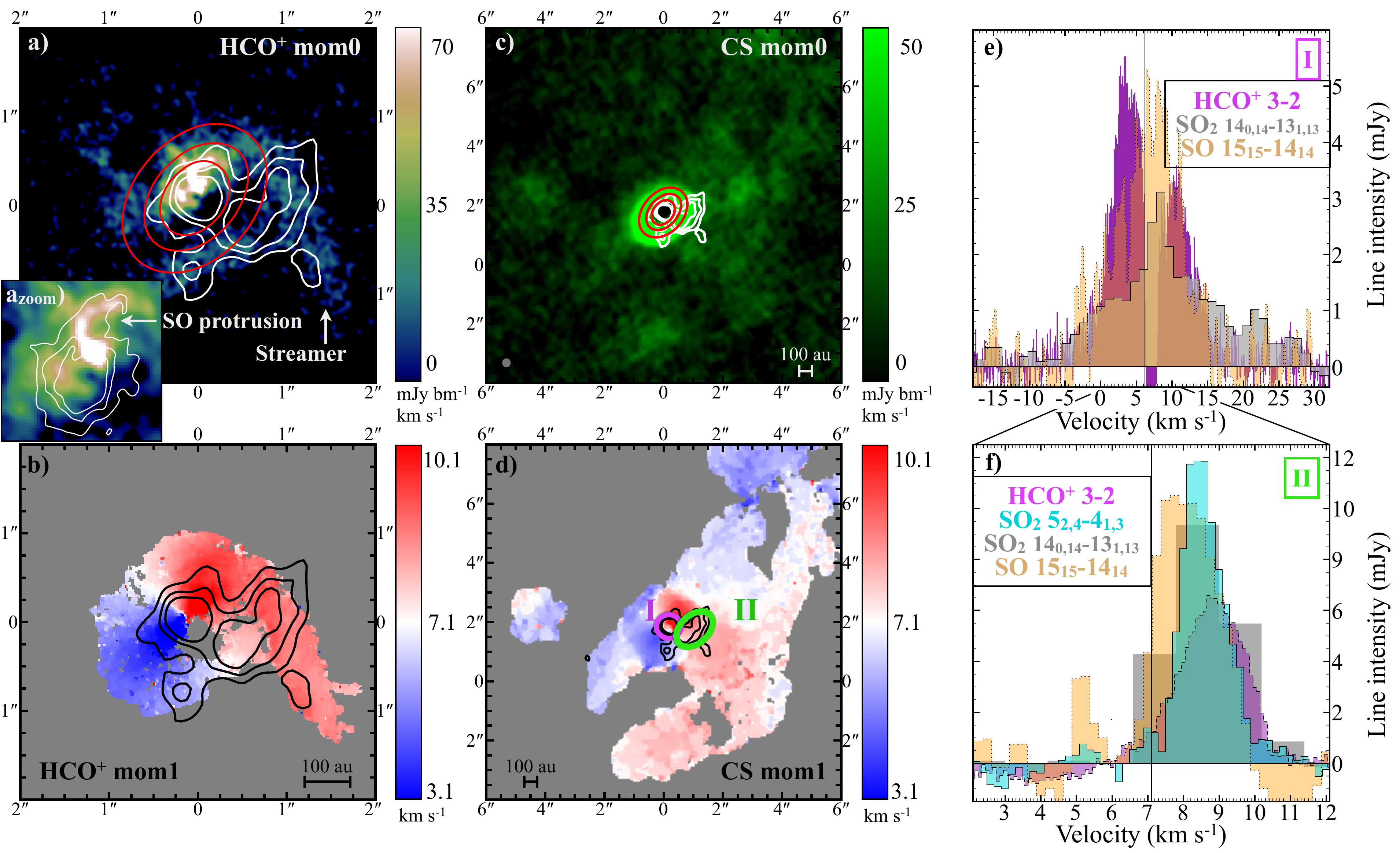} 
     \caption{Imagery of HL Tau. (a) Moment 0 of the HCO$^{+}$ 3$-$2 line \citep{Yen2019}. a$_{\rm zoom}$): zoom onto the central region of (a). (b) Moment 1 of the HCO$^{+}$ 3$-$2 line. (c) Moment 0 of the CS 5$-$4 line. (d) Moment 1 of the CS 5$-$4 line. (e) Integrated spectra of HCO$^{+}$, SO$_2$ $14_{0,14}-13_{1,13}$, and SO $15_{15}-14_{14}$ from the region labelled I in (d). (f) Integrated spectra of HCO$^{+}$, SO$_2$ $14_{0,14}-13_{1,13}$ and $5_{2,4}-4_{1,3}$ from the region labeled II in (b). In each map, red and white (or black) contours indicate the continuum and {SO$_2$ $14_{0,14}-13_{1,13}$} line emission as in Fig.\,\ref{Imagery_SO2}, except contours from (a$_{\rm zoom}$) that indicate the SO $15_{15}-14_{14}$ line emission. The arrow points to the inner part of the HCO$^{+}$ {streamer} visible {as a SO protrusion}. North is up, east is left.} 
 \label{Imagery_HLTau}
 \end{figure*}

\subsubsection{HL Tau} \label{HLTau}
The CO emission from HL Tau is dominated by the prominent outflow cones \citep[see][]{ALMA2015}. Here, we make use of the HCO$^{+}$ 3$-$2 maps described by \citet{Yen2019} and shown in Fig.\,\ref{Imagery_HLTau}a and \ref{Imagery_HLTau}b. The comparison with the SO and SO$_2$ detection described in Sect.\,\ref{Morphology} highlights that the inflowing spiral described by \citet{Yen2019} {and designated as a streamer in our figures} reconnects with the disk at the position where the outer component of the SO and SO$_2$ is detected. The SO$_2$ map even shows a weak protrusion that is spatially coincident with the HCO$^{+}$ {streamer}. A zoom onto the inner disk region (see Fig.\,\ref{Imagery_HLTau}a$\rm _{zoom}$) highlights the spatial consistency of the HCO$^{+}$ and SO $15_{15}-14_{14}$ line emission. More precisely, the inner part of the HCO$^{+}$ {streamer} corresponds to the northern protrusion of the central SO component shown in Fig.\,\ref{Imagery_SO2}.

On a larger scale, relatively bright, diffuse CS emission is detected to the west of the star (Fig.\,\ref{Imagery_HLTau}c). Even though there is no morphologically defined structure like the HCO$^{+}$ {streamer}, the velocity-weighted map {clipped at 5$\sigma$} (Fig.\,\ref{Imagery_HLTau}d) indicates that the CS to the SW has a peculiar, redshifted velocity comparable to that of the HCO$^{+}$ {streamer}. {When we apply an analytic streamline model to the HCO$^{+}$ and CS data together, we indeed find that both are consistent with an infalling streamer landing where the SO emission extends to the west (see Appendix \ref{App_HLTau}).}

The integrated spectra of the SO and SO$_2$ lines reveal that the emission from the intersection of disk and {streamer} (region II in \ref{Imagery_HLTau}d) has the same velocity pattern of the local disk traced by the HCO$^{+}$ emission (Fig.\,\ref{Imagery_HLTau}f) while the central component (region I) is very broad ($\sim$35 km s$^{-1}$, thus matching the tail of HCO$^{+}$) and approximately peaks at {systemic velocity ($V_{\rm sys} = +7.1$ km s$^{-1}$, \citealt{Garufi2021})} (Fig.\,\ref{Imagery_HLTau}e).

\begin{figure*}
  \centering
 \includegraphics[width=18cm]{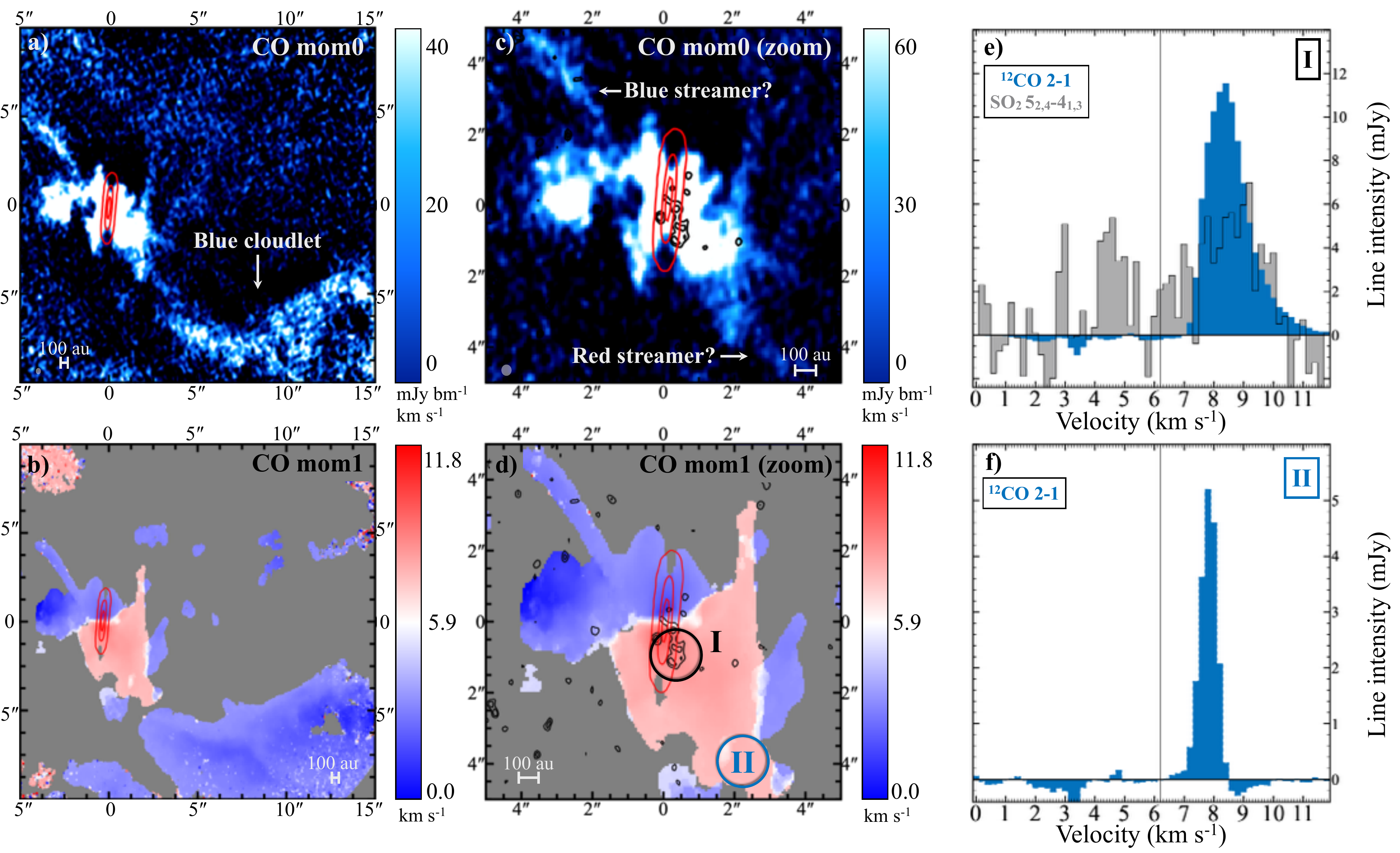} 
     \caption{Imagery of IRAS 04302+2247. (a) Moment 0 of the $^{12}$CO 2$-$1 line. b) Moment 1 of the $^{12}$CO 2$-$1 line. (c) Same as (a) on smaller scale. (d) Same as (b) on smaller scale. (e) Integrated spectra of CO and SO$_2$ 5$_{2,4}-4_{1,3}$ from the region labeled I in (d). (f) Integrated spectra of CO from the region labeled II in (d). In each map, red and black contours indicate the continuum and line emission as in Fig.\,\ref{Imagery_SO2}. North is up, east is left.
     } 
     
 \label{Imagery_IRAS}
 \end{figure*}

\subsubsection{IRAS 04302+2247} \label{IRAS04302}
The $^{12}$CO $2-1$ map of IRAS04302 reveals a complex environment with the presence of several arms. Any result on the global structure is biased by the absence of detectable signal from 3 to 7 km s$^{-1}$, {that is, for velocities close to the systemic velocity ($V_{\rm sys} = +5.6$ km s$^{-1}$, \citealt{Podio2020})}, because of the absorption by line-of-sight material. A very extended blueshifted cloudlet is seen at large scale to the SW of the source (see Figs.\,\ref{Imagery_IRAS}a and \ref{Imagery_IRAS}b). Closer to the source, the two most significant structures are the red and blue {streamers} (Figs.\,\ref{Imagery_IRAS}c and \ref{Imagery_IRAS}d) that seem to be the mirror opposites of each other. {From applying the analytic streamline solutions to these {structures} (Appendix \ref{App_IRAS}), we in fact find that the southern, redshifted {structure is consistent with a streamer and} intersects the disk} where the SO$_2$ signal is detected (see Sect.\,\ref{Morphology}), {while the northern, blueshifted structure is well described by the streamer model at small radii only}. Any connection between {the redshifted streamer} and the cloudlet remains speculative.

The integrated spectrum of Fig.\,\ref{Imagery_IRAS}e illustrates that the SO$_2$ signal is broad in velocity ($\sim5$ km s$^{-1}$) and slightly blueshifted. However, the comparison with CO reveals that it is not at the local disk velocity and that it is closer to that of the red {streamer} (Fig.\,\ref{Imagery_IRAS}f) although clearly broader.

\subsubsection{T Tau} \label{TTau}
\begin{figure*}
  \centering
 \includegraphics[width=18cm]{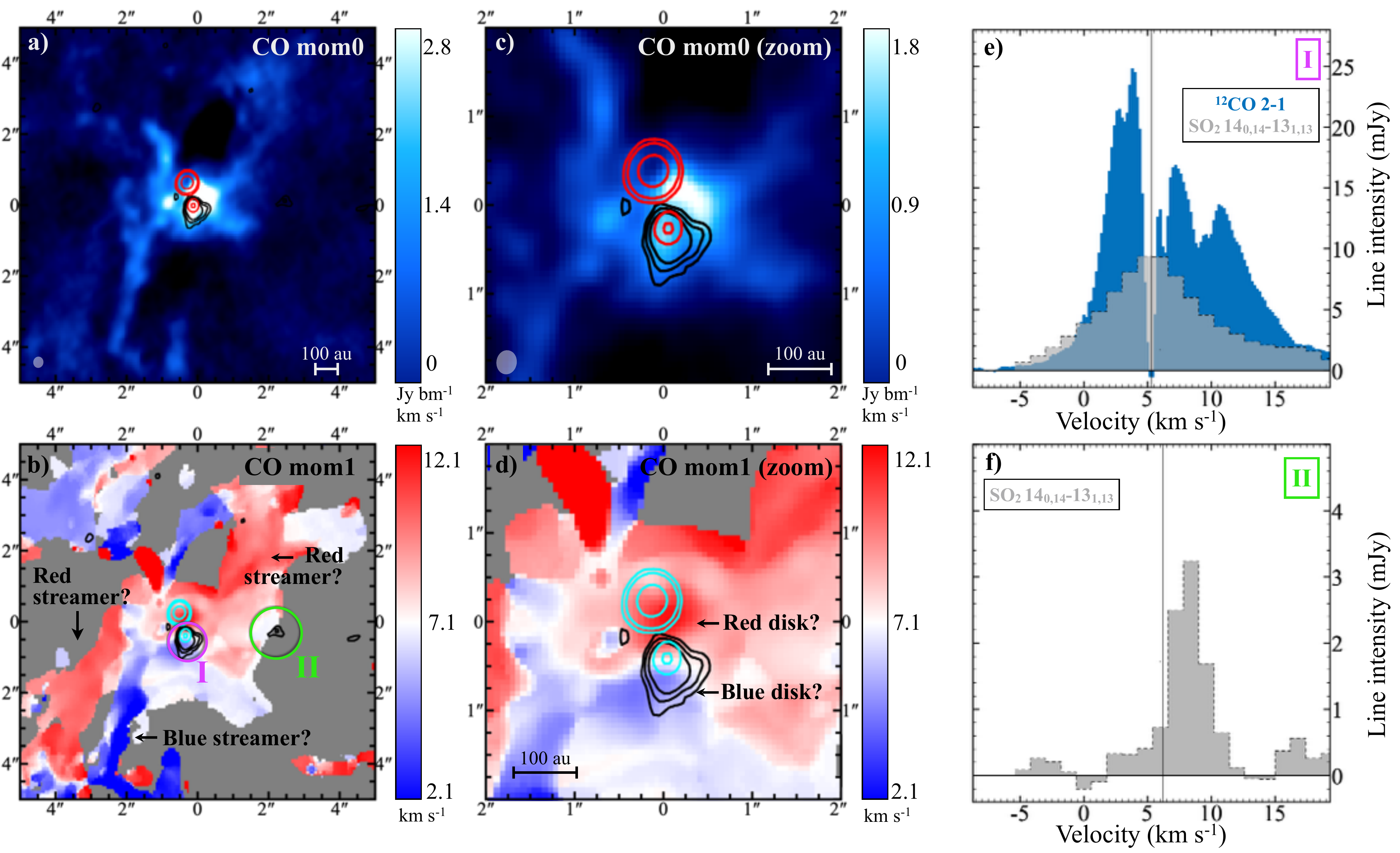} 
     \caption{Imagery of T Tau. (a) Moment 0 of the $^{12}$CO 2$-$1 line. (d) Moment 1 of the $^{12}$CO 2$-$1 line. (c) Same as (a) on a smaller scale. (d) Same as (b) on a smaller scale. (e) Integrated spectra of CO and SO$_2$ $14_{0,14}-13_{1,13}$ from the region labeled I in (b). (f) Integrated spectra of SO$_2$ $14_{0,14}-13_{1,13}$ from the region labeled II in (b). In each panel, red (or cyan) and black contours indicate the continuum and line emission as in Fig.\,\ref{Imagery_SO2}. North is up, east is left.} 
 \label{Imagery_TTau}
 \end{figure*}

The environment around the T Tau stellar system is notoriously very crowded \citep[see e.g.,][]{Duchene2005, Kasper2020}. The ALMA-DOT maps of the CS, CN, and H$_2$CO lines show multiple structures at a large scale \citep{Garufi2021} that are not in direct relation with T Tau N or S, and we will study these independently. In this work, we only examine the $^{12}$CO map because the emission peak from this line lies in proximity to T Tau S (see Fig.\,\ref{Imagery_TTau}a). In particular, the strongest integrated flux is detected NW of T Tau S, and appears to correspond to a bright bow-shaped feature seen in reflected light \citep{Kasper2016}. The absence of any signal centered on T Tau N could be explained by the nearly face-on geometry of its disk. {Generally speaking, such a} geometry only yields signal close to the rest-frame velocity and this is barely detectable in our maps because of the absorption by the large-scale material \citep{Garufi2021}.

The moment-1 map highlights the existence of multiple blue and red {streamers} that approximately run from SE to NW, or vice versa (see Fig.\,\ref{Imagery_TTau}b). {The streamline analysis of Appendix \ref{App_streamers} does not include these streamers because of the nearly face-on disk geometry of T Tau.} On a smaller scale {of a few tens of astronomical units (au)} (see Fig.\,\ref{Imagery_TTau}d), the CO velocity pattern around T Tau S exhibits what is expected from an inclined disk with position angle pointing north, and has a blue disk component to the south and a red disk component to the north. If the observed pattern were genuine disk emission, then the systemic velocity $V_{\rm sys}$ of the T Tau S binary would be around 7$-$8 km s$^{-1}$.

The SO$_2$ signal from T Tau S {extracted from region I of Fig.\,\ref{Imagery_TTau}b} is, similarly to the CO, very broad ($\sim25$ km s$^{-1}$, see Fig.\,\ref{Imagery_TTau}e) and centered on $\sim$5 km s$^{-1}$. This velocity corresponds to the main CO absorption feature, suggesting that it is the rest frame of the large-scale material. It does however differ from the aforementioned $V_{\rm sys}$ of T Tau S at 7$-$8 km s$^{-1}$ constrained by the CO pattern around the source. On the other hand, the signal from the western knot (region II in Fig.\,\ref{Imagery_TTau}b) is only detected in a narrow interval around 8$-$9 km s$^{-1}$ (see Fig.\,\ref{Imagery_TTau}f).

\subsection{HL Tau disk model}

\begin{figure}
  \centering
 \includegraphics[width=9cm]{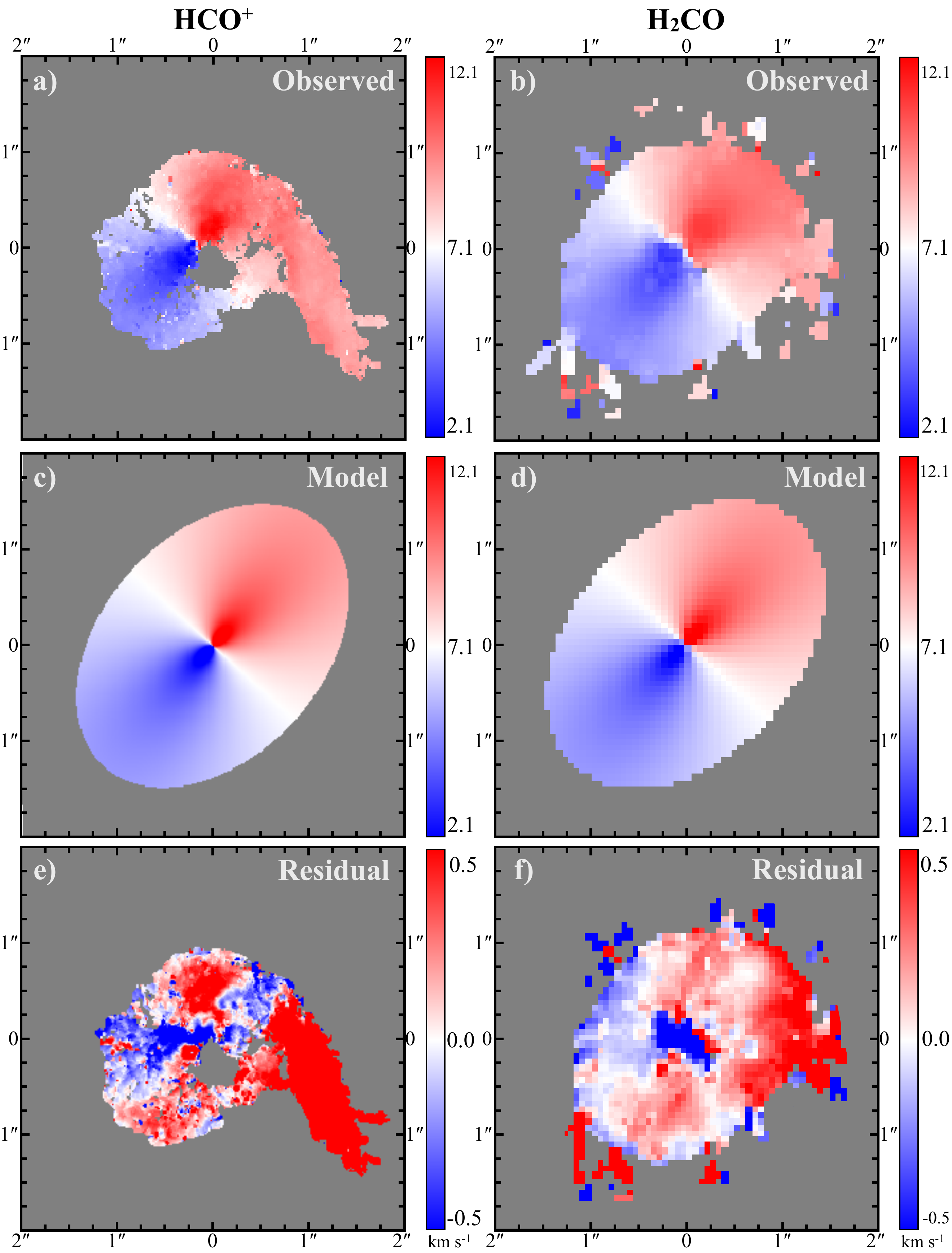} 
     \caption{HL Tau disk velocity structure analysis. {A Keplerian disk model (c and d) is subtracted from the observed HCO$^+$ and H$_2$CO moment-1 maps \citep[a and b, from][]{Yen2019, Garufi2021},} yielding the velocity residuals of (e) and (f).  For both HCO$^{+}$ and H$_2$CO, outside of the disk radius the $v_{\rm sys}$ is subtracted.}
 \label{Model_HLTau}
 \end{figure}

To determine what impact the {streamer} might have on the HL Tau disk kinematics, the predicted Keplerian velocity of each pixel is calculated assuming a geometrically flat disk \citep[][]{Pineda2019}. We adopted a stellar mass of 2.1 M$_{\rm \odot}$ \citep{Yen2019}, a distance of 147.3 pc \citep{Galli2018}, a systemic velocity $V_{\rm sys}$ of 7.1 km s$^{-1}$, a position angle of 135$^{\circ}$, a disk inclination angle of 35$^{\circ}$, and a disk outer radius of 250 au  \citep{Garufi2021}. We convolved the velocity model with the beam of the HCO$^{+}$ {(0.1\arcsec$\times$0.09\arcsec)} and H$_2$CO ALMA-DOT {(0.31\arcsec$\times$0.26\arcsec)} observations. Within the disk radius, we subtracted the Keplerian velocity model, while outside of the disk outer radius we subtracted the systemic $V_{\rm sys}$, ensuring a smooth connection at the boundary between the disk and spiral structure \citep{Akiyama2019}. The disk model and residuals after subtracting the Keplerian model from the HCO$^{+}$ and H$_2$CO moment-1 maps are shown in Fig.\,\ref{Model_HLTau}. The HCO$^{+}$ and H$_2$CO velocity patterns after subtracting the expected rotation are very similar. Both reveal redshifted material to the south, north, and in the HCO$^{+}$ {streamer} region to the west, as well as blueshifted material or material close to the $V_{\rm sys}$ to the east and northwest. These results are discussed in Sect.\,\ref{HLTau_kinematics}.

\section{Discussion} \label{Discussion}
The results of Sect.\,\ref{Results} can be summarized as follows:
\begin{itemize}
    \item {The spatial distribution of SO and SO$_2$ emission detected in the Class I/II disks of DG Tau, HL Tau, IRAS04302, and T Tau does not follow the dust and gas distribution in the disk, that is, it does not probe a symmetric radial and/or vertical disk region as observed in the 1.3 mm continuum or in other molecular tracers such as H$_2$CO and CS (see, e.g., the overview of the continuum and molecular emission in the ALMA-DOT disks in Figure 1 by \citealt{Garufi2021}). }
    
    \item {In the cases of DG Tau and HL Tau, the SO and SO$_2$ emission is not located along the jet direction and/or on the outflow cavities as mapped by previous authors in both atomic \citep[e.g., ][]{Mundt1990,Eisloffel1998} and molecular \citep[e.g., ][]{ALMA2015,Guedel2018} lines.
    } 
    
    \item {In all four disks, we detect SO$_2$ (as well as SO for DG Tau and HL Tau) emission from only one side of the disk, from a compact region which is displaced by $20$ to $190$ au with respect to the dust  continuum peak. This emission is detected at the intersection between the disk and some} streamers observed in CO, HCO$^{+}$, or CS. 
    
    \item The velocities of the {SO and SO$_2$} emission {from these outer disk regions} are comparable with those of the streamers. Their {spectral width} is small (5$-$7 km s$^{-1}$). 
    
    \item {In the case of HL Tau, IRAS04302, and T Tau, we also detect bright emission centered on the continuum peak from the inner $20$ au disk region. This emission is detected only in the high-excitation SO$_2$ line ($E_{\rm up} = 93$ K) and SO lines ($E_{\rm up} = 253-261$ K, only for HL Tau).} 
    
    \item The {SO and SO$_2$} emission {from the inner disk is} centered at the systemic velocity of the source, and {covers a large velocity interval} (25$-$35 km s$^{-1}$). {In the case of HL Tau, the moment-1 map shows that the SO gas has a velocity pattern in agreement with the disk Keplerian rotation pattern observed in other gas tracers.} 
    
    \item The gas temperature associated with the {SO and SO$_2$ emission from the outer disk of HL Tau is $\sim 60$ K, while it is $>75$ K in the inner region of IRAS04302 and $>350$ K in the inner disk of HL Tau.}

\end{itemize}

In this section, we discuss these results, focusing on the origin of the observed SO and SO$_2$ emission in the context of the observed connection between disk and environment.

\subsection{Inflow or outflow} \label{In_outflow}
\begin{figure*}
  \centering
 \includegraphics[width=18cm]{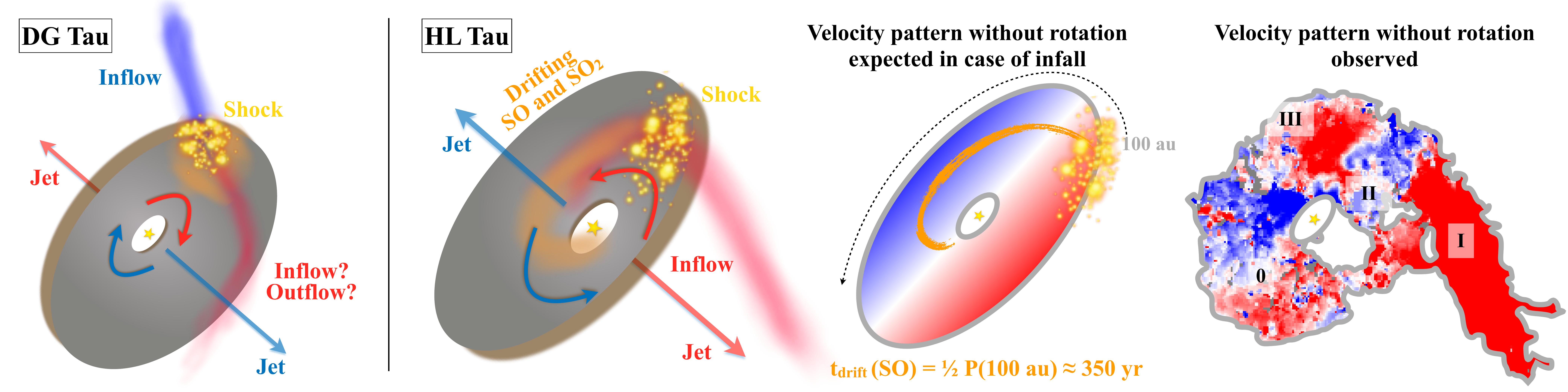} 
     \caption{Illustrative sketch of the environment of DG Tau and HL Tau. The third panel shows the velocity pattern expected in case of uniform infall after rotation subtraction. The fourth panel shows the observed HCO$^{+}$ velocity pattern after rotation subtraction (see Fig.\,\ref{Model_HLTau}).}. 
 \label{Sketch}
 \end{figure*}

To investigate the origin of the observed SO and SO$_2$ emission, we must first determine whether the observed structures around the disk are inflowing or outflowing. The northern streamer observed in CS at large scale around DG Tau (see Fig.\,\ref{Imagery_DGTau}) is most likely the same structure yielding the CO blue component. \citet{Guedel2018} concluded that this component is outflowing material. However, the detection of the CS northern streamer suggests that we are observing an inflowing structure, {which is confirmed by the analytic infalling streamline model in Appendix \ref{App_DGTau}}. This is appreciable from the sketch of Fig.\,\ref{Sketch}. 
On the other hand, the {southern streamer} (see Fig.\,\ref{Imagery_DGTau}) could be either inflowing or outflowing, {because it may be associated with the northern streamer piercing through the disk and about to turn back towards the young stellar object}. In both cases, the angle of incidence with the disk should be smaller than the disk inclination \citep[35\degree,][]{Podio2020b} to enable a receding velocity. It is therefore tempting to propose that the {southern streamer} is the continuation of the northern {streamer}. In that putative case, a significant amount of material should not efficiently accrete onto the disk but should instead deflect by a considerable angle driven by the disk rotation. In other words, redshifted disk material would cause blueshifted accreting material to become redshifted outflowing material.

The interpretation of the {streamer} in HL Tau (see Fig.\,\ref{Imagery_HLTau}) is rather intuitive. Inflowing material is impacting the disk on the redshifted side with a coherent direction (NE, see the sketch of Fig.\,\ref{Sketch}). This is the same conclusion as that drawn by \citet{Yen2019}, {and is supported by our streamline modeling in Appendix \ref{App_HLTau}}. The kinematics of the material around HL Tau is further discussed in Sect.\,\ref{HLTau_kinematics}.

{Confirming the inflow of material in the red and blue streamers of IRAS04302 (see Fig.\,\ref{Imagery_IRAS}) is more complicated, because these streamers lie along the outflow cavity}. This source is also known as the butterfly star because of its prominent bipolar morphology observed by \citet{Lucas1997} and \citet{Padgett1999}, which is recurrently interpreted as outflow cavities. These structures are  probably\ not evident in our CO maps because their expanding projected velocity is nearly zero given the disk edge-on geometry, and any material at the source systemic velocity is not observable (see Sect.\,\ref{IRAS04302}). In principle, the intrinsic outflow rotation may add a velocity component that enables some material to emit at detectable velocities. However, in DG Tau B and HL Tau \citep{Garufi2021, ALMA2015}, the outflow rotation only introduces a $\pm1$ km s$^{-1}$ contribution to the net projected velocity and therefore does not change this view. Geometrically, the blue and red {streamers} {of IRAS04302} could be {instead} part of the outflow cavities but this would require a mechanism that induces a large rotation velocity on only one side of the outflow cavities; {furthermore, our analytic streamline modeling supports the infall scenario for both streamers (Appendix \ref{App_IRAS})}. Also, the {structures in question} do not seem to originate from the star (see Fig.\,\ref{Imagery_IRAS}) and therefore it is more likely that we are observing accreting features similar to the cases of DG Tau and HL Tau.

Finally, the case of T Tau is very complex. The main emission is clearly associated with T Tau S and is extended toward the south (see Fig.\,\ref{Imagery_TTau}). The Subaru and ALMA observations by \citet{Yang2018} and \citet{Manara2019} suggest that the circumbinary disk of T Tau S is approximately oriented north$-$south, in agreement with our speculations outlined in Sect.\,\ref{TTau}. In this scenario, the rest-frame velocity of the disk would be 7$-$8 km s$^{-1}$, and the SO$_2$ emission peaking at 5 km s$^{-1}$ (see Fig.\,\ref{Imagery_TTau}e) would be blueshifted, in analogy with the velocity of the disk toward the south. This may suggest that the SO$_2$ signal is disk emission. Alternatively, it could be associated with one of the several outflows inferred in this system \citep{Herbst1997, Herbst2007}. A blueshifted, wide-angle outflow to the SE is believed to originate from T Tau S \citep{Kasper2016}. It is tempting to associate the prominent CO blueshifted arm from Fig.\,\ref{Imagery_TTau}b to this outflow but its origin does not seem to be T Tau S. On the other hand, the series of SO$_2$ knots visible from Figs.\,\ref{Imagery_SO2} and \ref{Imagery_TTau} is likely associated with the east--west outflow from T Tau N \citep{Bohm1994}, and in particular the western knots lie in the direction of the prominent bow shock detected in the NIR \citep{Kasper2020}, but at a larger separation.

\subsection{Origin of the SO and SO$_2$ emission} \label{SO_discussion}

The localized nature of the SO and SO$_2$ emission at the intersection between disk and molecular streamers around DG Tau and HL Tau indicates a confined increase of column densities. The most likely origin of such a {localized} increase is shocks {due to} the infalling streamers {impacting the disk, which could cause a release of SO and SO$_2$ molecules from the sputtered dust grain mantles. In fact, any thermal release of SO and SO$_2$ in the disk layer at temperatures above their evaporation temperature would instead result in an azimuthally symmetric distribution of emission}. These shocks {would be} reminiscent of those associated with Class 0 sources with a denser protostellar envelope. In these earlier objects, such as L1527 or B335 \citep[e.g.,][]{Sakai2014,Sakai2017,Oya2016,Oya2017,Imai2019}, slow shocks (around 1 km s$^{-1}$) occur at the transition zone of the infalling and rotating envelope with the protostellar accretion disk, {causing an enhancement of the SO emission in a ring-like structure}. In the ALMA-DOT sample studied in this work, the envelope is largely dissipated. Therefore, a ``well-behaved'' symmetric ring of shocked material in front of the centrifugal barrier is not expected. Instead, shocks are located along the late infalling streamers still feeding the young stellar objects. The temperatures derived {in the SO- and SO$_2$-emitting region} ($\ge$ 60 K, see Sect.\,\ref{Temperatures}) are in agreement with emission from shocked material. 

As discussed in Sect.\,\ref{Emission_morphology}, HL Tau also shows a bright SO and SO$_2$ component centered on the star. The emission from these regions may also be related to the action of inner jets and outflow \citep[e.g.,][]{Podio2021} or to the innermost regions of disk and envelope \citep[e.g.,][]{Harsono2021} where dust mantles can sublimate, thus resembling a hot-corino chemistry. \citet{Booth2021} also showed an asymmetric SO and SO$_2$ emission co-spatial with a dust crescent where the molecular enhancement would be related to the sublimation of ices at the edge of a dust cavity at a separation of $\sim$ 50 au. However, in the case of HL Tau presented here, a further explanation is possible. The central SO component in question is clearly protruded toward the shocked region at $\sim$100 au from the star following the inward motion of the HCO$^{+}$ {streamer}  (see Figs.\,\ref{Imagery_SO2} and \ref{Imagery_HLTau}). This suggests that the SO and SO$_2$ molecules released in the shock at $\sim$100 au spiral toward the star before any chemical reprocessing occurs. The drifting timescale can be coarsely estimated from the morphology of the SO spiral, covering approximately 180$\degree$ (see the sketch in Fig.\,\ref{Sketch}). The orbital period at 100 au from a 2.1 $\rm M_{\odot}$ star \citep{Yen2019} is $\sim$700 yr. Thus, the spiraling material drifts from 100 au to the star in less than 350 yr, corresponding to a velocity of 1$-$2 km s$^{-1}$. These velocities are those of the SO and SO$_2$ emission at the shock location and of the HCO$^{+}$ accreting {streamer} (see Fig.\,\ref{Imagery_HLTau}f), giving support to the proposed explanation. A timescale of a few hundred years is also shorter than the expected timescale needed for the SO and SO$_2$ abundances to significantly decrease after the occurrence of a shock. This value depends on a number of parameters, but, following for example\,\citet{Pineau1993}, \citet{Charnley1997}, and \citet{Taquet2019},
can be roughly estimated as more than 10$^3$ yr, before their abundances are decreased by one order of magnitude.

\subsection{Influence of the streamer on disk kinematics in HL Tau} \label{HLTau_kinematics}

 The dust rings of HL Tau lie in the geometrically thin disk midplane \citep{ALMA2015} and do not show signs of disturbance in the regions of the disk that spatially overlap with the streamer. This indicates that the streamer spirals into the central region without passing through the disk midplane while the gas kinematics {in the molecular layer} are clearly affected by the accreting material (Sect.\,\ref{HLTau}). {This finding is consistent with models of disk accretion occurring at the upper layer of the disk surface \citep{Bai2016, Riols2020}, and is among the first direct detections of surface accretion at large radii ($50-100$ au), while recent evidence was provided for the inner regions of the disk \citep[$<10$ au,][]{Najita2021}.} 

 To interpret the observed velocity pattern of 
HCO$^{+}$ and H$_2$CO after removing the expected Keplerian rotation (see Fig.\,\ref{Model_HLTau}), we must first consider that a uniform infalling or inwardly drifting component should be reflected in the residual maps. For the geometry of the disk of HL Tau (with the far side of the disk to the NE; see sketch of Fig.\,\ref{Sketch}), a blueshifted infalling component should appear to the NE and a redshifted infalling component to the SW. The SE portion of the disk (region 0 in Fig.\,\ref{Sketch}), which does not significantly overlap with the accreting material, shows this behavior. 

Conversely, the NW portion of the disk shows the expected redshifted component associated with the {streamer} (region I), as well as a blueshifted component near the intersection with the disk (region II) and a red component where the SO and SO$_2$ molecules are thought to drift inward (region III). A possible interpretation of the blueshifted material in region II is that the accreting material heats the local disk, resulting in a puffed-up uppermost disk layer that is probed in our maps by its approaching velocity. After that, the cooling streamer material moves closer to the disk midplane and pushes the denser disk material as it orbits, adding a receding velocity component to the net velocity pattern in region III.  After this point, the streamer material has drifted inwards to the central protostar (Sect.\,\ref{SO_discussion}) without ever crossing the disk midplane, leaving the SE portion of the disk (region 0) undisturbed from the expected rotation and uniform infall pattern. This proposed scenario could be tested with modeling.

\section{Conclusions} \label{Conclusions}
Increasing attention from the planet-formation community is being placed on the interaction between planet-forming disks and their surrounding medium. The discovery of several streamers feeding the disk \citep[e.g.,][]{Akiyama2019, Yen2019, Pineda2020} offers the opportunity to study how disk accretion proceeds at late stages, when planet formation is possibly already ongoing.

In this study, we show some {extended} structures resembling the aforementioned streamers around four sources that are still partly embedded in their natal cloud (Class I or early Class II objects). More importantly, we revealed SO and SO$_2$ emission that appears to correspond to the intersection between disk and streamers. Two of the four cases in question, DG Tau and HL Tau, are clear cases of inflowing material impacting the circumstellar disk and inducing a shock that is traced by {emission of SO and SO$_2$  discretely localized on the region where the streamers connect to the disk}. Unlike younger Class 0 sources, such shocks are confined to specific disk regions {\citep[e.g.,][]{Lee2019}}. 

While in DG Tau the SO and SO$_2$ emission is only detected at 50 au from the star, in HL Tau their emission is probed from the outermost disk region, where the shock occurs, down to the innermost disk region. Our interpretation of the inner component of the emission is that the SO and SO$_2$ molecules released in the shock in the outskirts of the disk  spiral toward the star in less than 350 years before any chemical process may occur. We also reveal that the disk kinematics is altered by the accreting material, as the disk shows a blueshifted component in proximity to the shock that can be ascribed to heated material being uplifted toward the observer.

{In IRAS 04302+2247, the SO$_2$ emission is much weaker but is also possibly associated with the physical intersection between the disk and the streamer. Finally,} T Tau is more ambiguous because of the complex environment. It is still possible that the SO$_2$ emission detected in this target originates from inflowing material, although the presence of prominent outflowing structures offers a valid alternative mechanism that could induce the observed shocks.

The possibility that {Class I and II} planet-forming disks in Taurus or other {relatively old} star-forming regions {are characterized by} accreting streamers {infalling onto the disk and shocking its material} is a realistic possibility that must be taken into account when analyzing the disk structure. 
{Our observations demonstrate that the circumstellar clouds and protoplanetary disks have complex, dynamic structures and that material is most likely continuously fed onto the disk, even in Class II sources, changing the disk temperature and density profiles, as well as altering its chemistry.  This addition of material onto the protoplanetary disk should be considered when modeling planet formation and calculating planetary mass budgets.  In addition, such accretion processes should also be considered in regards to the stability of the protoplanetary disk.}
The prototypical example described in this work is also the most studied, namely HL Tau, where the presence of a shock that is possibly altering the whole disk kinematics has not been considered before. This paper further shows that S-bearing molecular species such as SO and SO$_2$ {may be used to probe such accretion shocks caused by late accretion events onto the disk} in Class I and II objects, thus extending a branch of investigation that has been profitable in younger, Class 0 sources.

\begin{acknowledgements}
     {We thank the referee for the constructive report.} We are also grateful to K.\,Rygl and C.\,Spingola for the help with the ALMA data and to H.-W.\,Yen for sharing their ALMA data. This paper makes use of the following ALMA data: ADS/JAO.ALMA\#2016.1.00846.S, 2017.1.01178.S, 2017.1.01562.S, and 2018.1.01037.S. ALMA is a partnership of ESO (representing its member states), NSF (USA) and NINS (Japan), together with NRC (Canada), MOST and ASIAA (Taiwan), and KASI (Republic of Korea), in cooperation with the Republic of Chile. The Joint ALMA Observatory is operated by ESO, AUI/NRAO and NAOJ. This work was supported by the PRIN-INAF 2016 "The Cradle of Life - GENESIS-SKA (General Conditions in Early Planetary Systems for the rise of life with SKA)", the project PRIN-INAF-MAIN-STREAM 2017 "Protoplanetary disks seen through the eyes of new-generation instruments",     the program PRIN-MIUR 2015 STARS in the CAOS - Simulation Tools for Astrochemical Reactivity and Spectroscopy in the Cyberinfrastructure for Astrochemical Organic Species (2015F59J3R, MIUR Ministero dell'Istruzione, dell'Universit\`{a}, della Ricerca e della Scuola Normale Superiore), the European Research Council (ERC) under the European Union's Horizon 2020 research and innovation programme, for the Project "The Dawn of Organic Chemistry" (DOC), Grant No 741002, the European MARIE SKLODOWSKA-CURIE ACTIONS under the European Union's Horizon 2020 research and innovation programme, for the Project "Astro-Chemistry Origins" (ACO), Grant No 811312, INAF/Frontiera (Fostering high ResolutiON Technology and Innovation for Exoplanets and Research in Astrophysics) through the "Progetti Premiali" funding scheme of the Italian Ministry of Education, University, and Research, as well as NSF grants AST- 1514670 and NASA NNX16AB48G, the Italian Ministero dell'Istruzione, Universit\`{a} e Ricerca through the grant Progetti Premiali 2012 - iALMA (CUP C52I13000140001), the Deutsche Forschungs-Gemeinschaft (DFG, German Research Foundation) - Ref no. FOR 2634/1 TE 1024/1-1, by the DFG cluster of excellence ORIGINS (www.origins-cluster.de), by Wallonie-Bruxelles International (Belgium) through its grant "Stage en Organisation Internationale", and by the European Union's Horizon2020 research and innovation programme under the Marie Sklodowska-Curie grant agreement No 823823 (RISE DUSTBUSTERS project) and from the European Research Council (ERC) via the ERC Synergy Grant {\em ECOGAL} (grant 855130).
     \end{acknowledgements}

\bibliographystyle{aa} 
\bibliography{MasterReference.bib} 

\begin{appendix}
\section{Rotational diagrams}
\label{sect:RD}

\begin{figure}
  \centering
 \includegraphics[width=9cm]{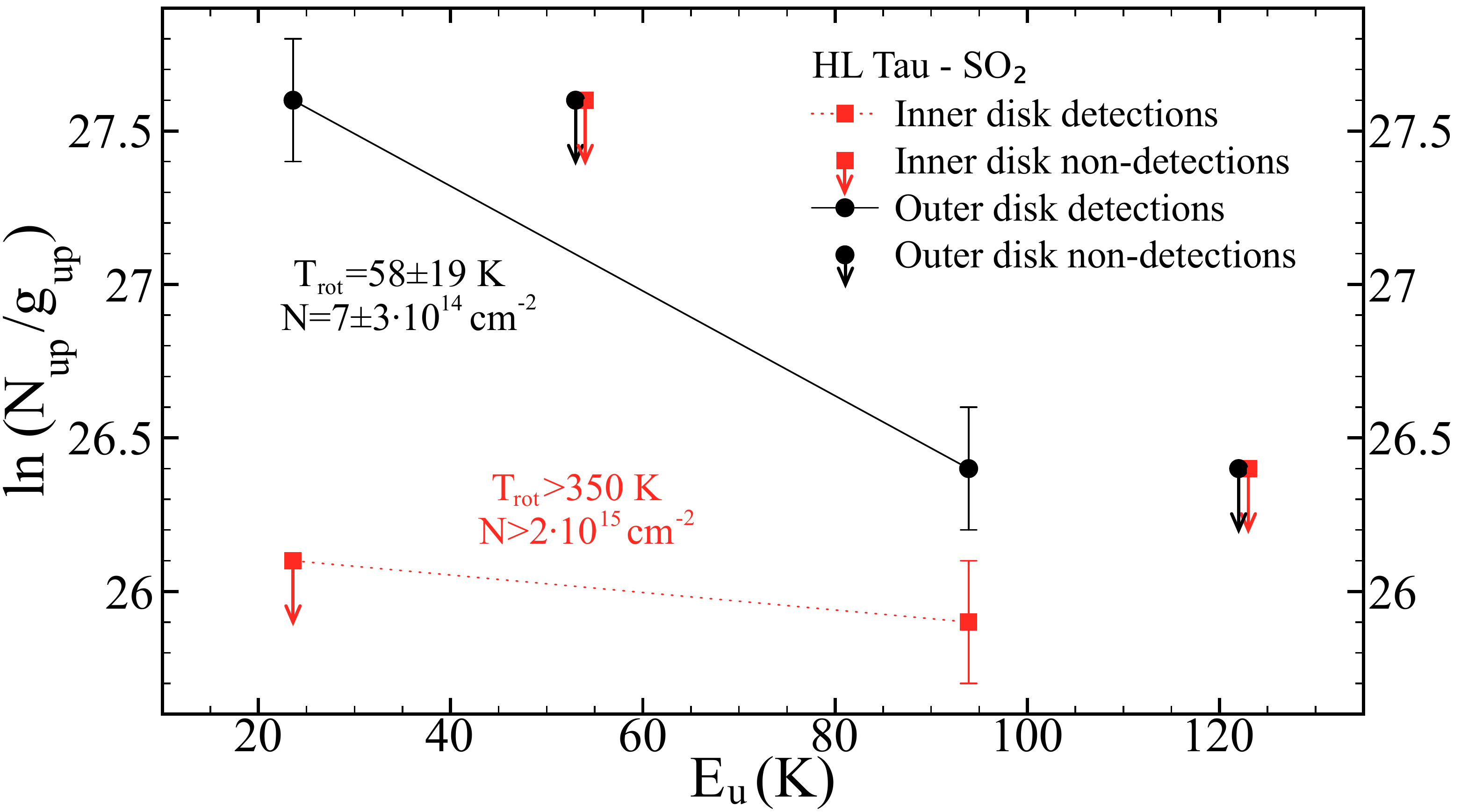}
  \includegraphics[width=9cm]{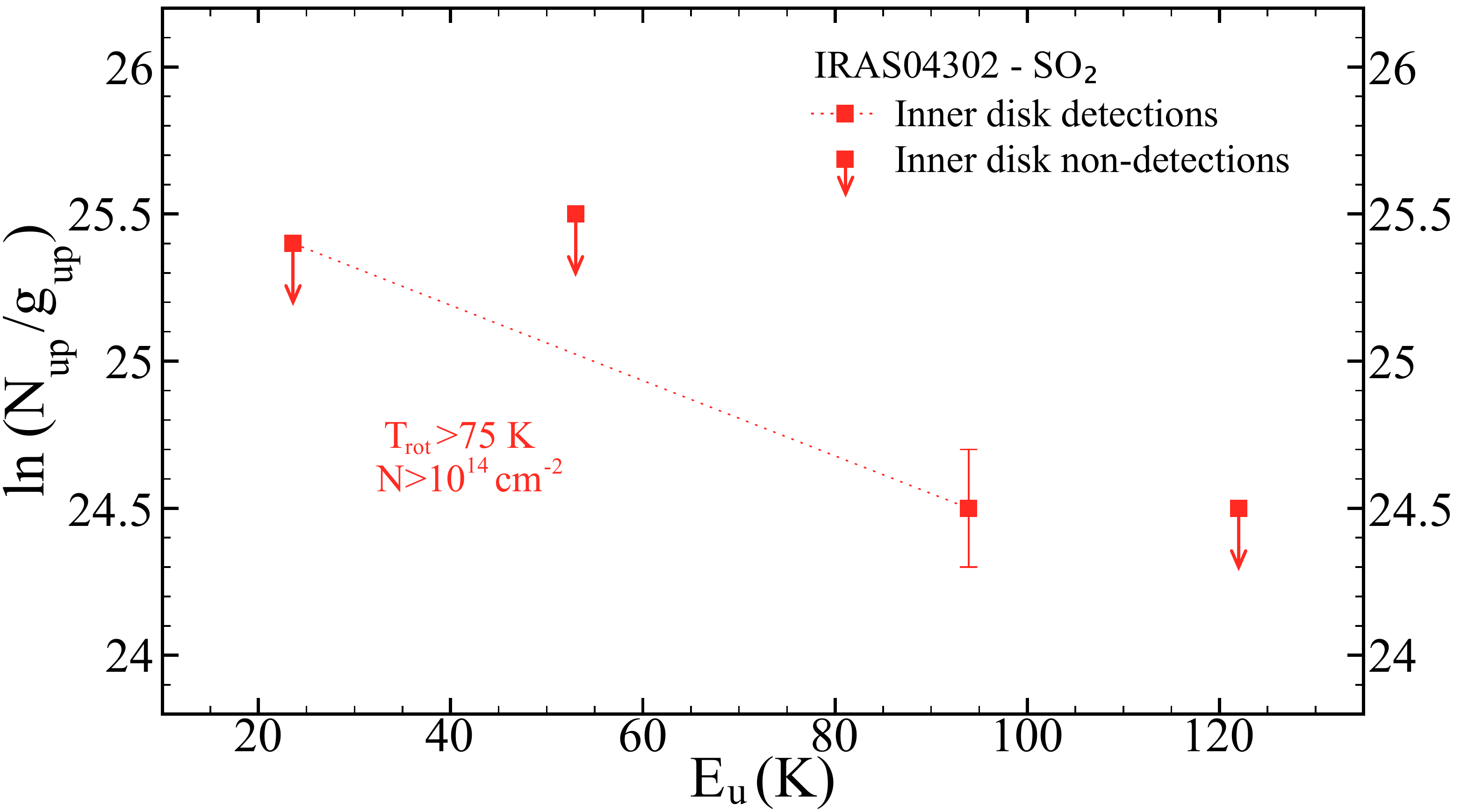}
   \includegraphics[width=9cm]{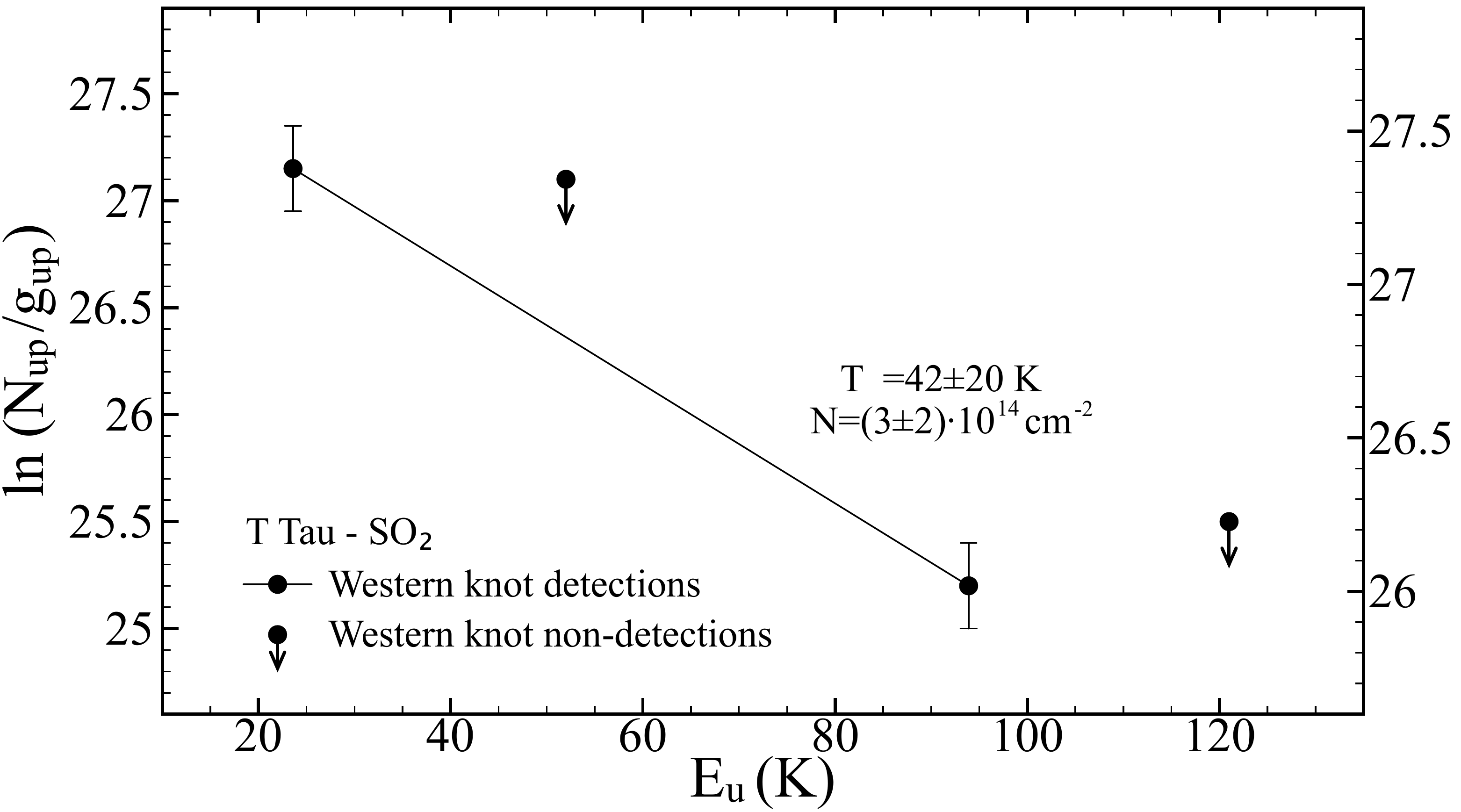}
     \caption{Rotational diagrams of the SO$_2$ lines detected for HL Tau (outer and inner disk), IRAS04302, and the knot located $2.5\arcsec$ west of T Tau S. The derived values of $T_{\rm rot}$ and $N$ are labeled. The upper limits for undetected SO$_2$ are overplotted to check for consistency with the obtained solution.}. 
 \label{RD}
 \end{figure}

{The rotational diagrams obtained for the SO$_2$ lines detected in HL Tau (outer and inner disk), IRAS04302, and the knot located $2.5\arcsec$ west of T Tau S, are shown in Figure \ref{RD}.}

\section{Confirming infall with streamline models} \label{App_streamers}
{The trajectory of material infalling towards a central mass through a streamer can be described by an analytic streamline model of material in a rotating and collapsing sphere \citep[see][]{Mendoza2009,Pineda2020}. This analysis constitutes a generalization of the Ulrich profile, which is often used to describe material in systems with both envelopes and disks \citep{Ulrich1976}. When possible, we manually and iteratively search for solutions of the streamline model which describe both the spatial and kinematic structure of the possible streamers. The T Tau data were not suitable for such an analysis because of the nearly face-on disk geometry, which makes it nearly impossible to kinematically discriminate between different models.}

{Several parameters of the streamline models were fixed for consistency with previous ALMA-DOT studies \citep{Garufi2021}: the central stellar mass ($M_{star}$), the central rest velocity of the system ($V_{lsr}$),  the disk inclination angle ($I.A._{disk}$), and the disk position angle ($P.A._{disk}$).
Parameters that were manually explored in the streamline modeling process include the initial streamline positional parameters of radius, polar angle, and azimuthal angle on the rotating sphere ($r_0$, $\theta_0$, $\phi_0$).  The dynamical parameters of the streamline initial angular velocity ($\Omega_0$) and initial infall velocity ($v_{r,0}$) were also explored. The model parameters that we successfully found to largely reproduce the spatial and kinematic structures are presented in Table \ref{tab:stream_params}, and a brief discussion of each of the streamers in DG Tau, HL Tau, and IRAS 04302+2247 are discussed in the following sections.}

\begin{table*}[ht]
\centering
\caption{Streamline model parameters}
\label{tab:stream_params}
\begin{tabular}{ccccc}
\hline
Parameter  & DG Tau & HL Tau & IRAS 04302+2247 (South)$^{1}$ & IRAS 04302+2247 (North)$^{1}$   \\ 
\hline
Fixed parameters: &&&& \\
$M_{star}$ & 0.3 M$_\odot$ & 2.1 M$_\odot$& 2.0 M$_\odot$ & 2.0 M$_\odot$ \\
$V_{lsr}$  & 6.2 \kms & 7.1 \kms & 5.9 \kms & 5.9 \kms \\
$I.A._{disk}$$^{2}$ & 35$^{\circ}$ & 47$^{\circ}$ & 81$^{\circ}$ &  81$^{\circ}$ \\
$P.A._{disk}$$^{3}$       & 135$^{\circ}$  & 138$^{\circ}$& 355$^{\circ}$ & 355$^{\circ}$  \\ 
\hline
Explored parameters: &&&& \\
$r_0$       & 450  au & 700 au & 1200 au & 1200 au \\
$\theta_0$ & 100$^{\circ}$  & 95$^{\circ}$ & 70$^{\circ}$& 120$^{\circ}$  \\
$\phi_0$   & 295$^{\circ}$  & 240$^{\circ}$ &  315$^{\circ}$ & 135$^{\circ}$\\
$\Omega_0$ &   $2.0 \times 10^{-12}$ s$^{-1}$ & $7.0 \times 10^{-12}$ s$^{-1}$ & $2.0 \times 10^{-12}$ s$^{-1}$  & $5.0 \times 10^{-13}$ s$^{-1}$\\
$v_{r,0}$   & 0.4 \kms  & 0.1 \kms & 2.1 \kms  & 4.3 \kms \\ \hline
\end{tabular}
 \tablefoot{{$^{1}$IRAS 04302+2247 (south) and IRAS 04302+2247 (north) respectively refer to the southern and northern streamers feeding the IRAS 04302+2247 disk. $^{2}$A disk with $I.A._{disk}$=0$^{\circ}$ inclination angle is face-on with respect to the observer.  $^{3}$A disk with $P.A._{disk}$=0$^{\circ}$ is oriented in the north--south direction, with $P.A._{disk}$ increasing counter-clockwise from the north.}}
\end{table*}

\subsection{DG Tau}\label{App_DGTau}
\begin{figure}
  \centering
 \includegraphics[width=9cm]{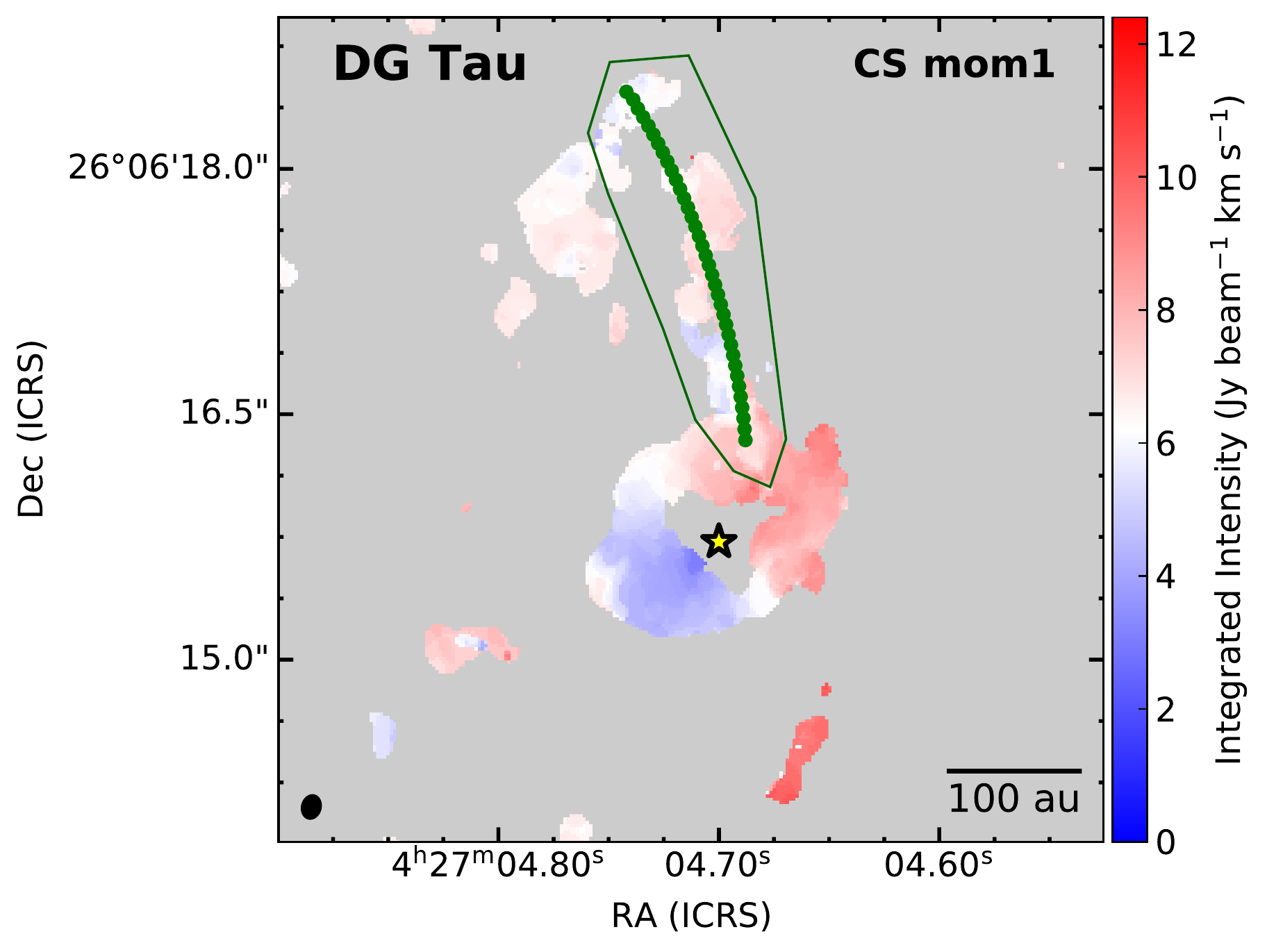}
 \includegraphics[width=9cm]{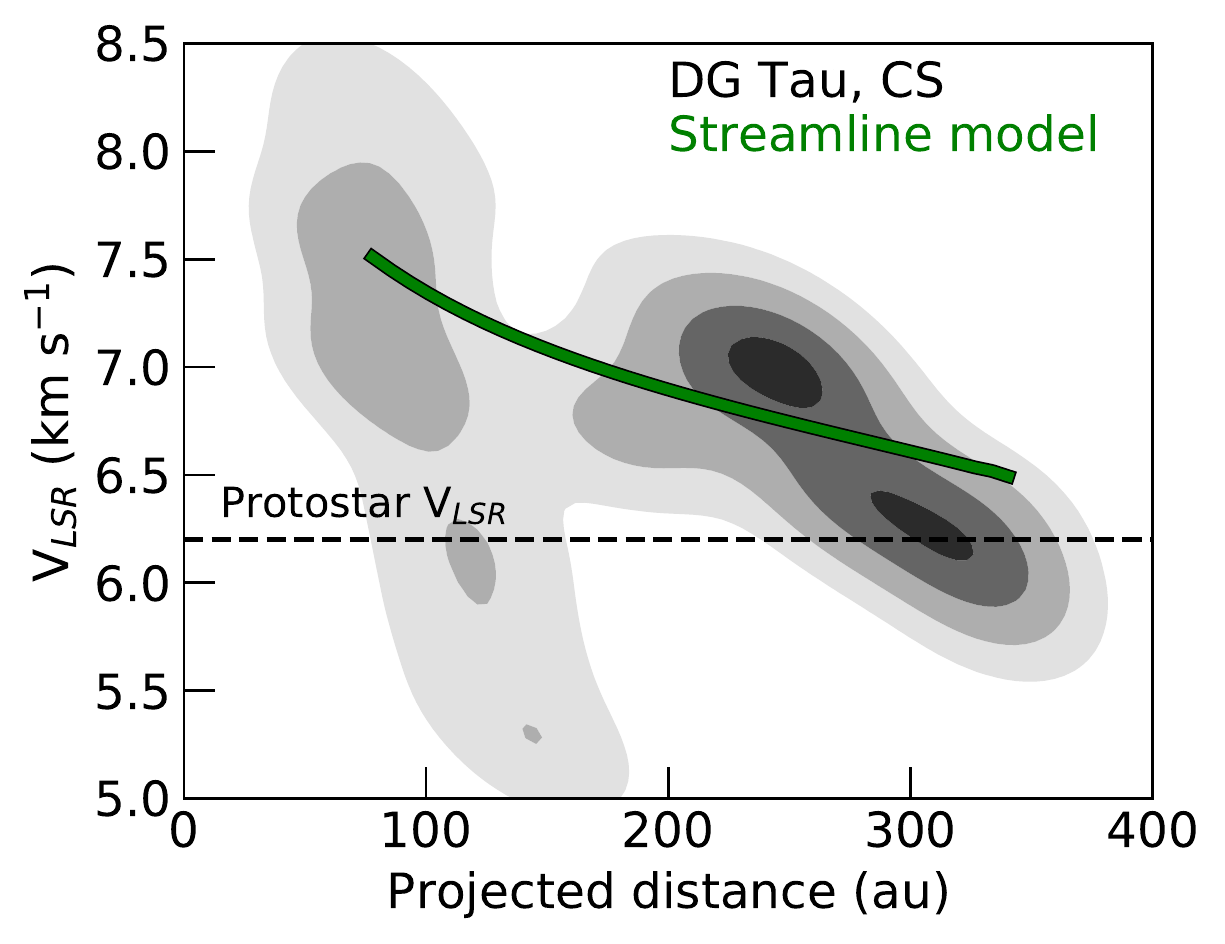}
     \caption{Infalling streamline model for the DG Tau streamer. Top: Streamline model, plotted in green, on top of the moment-1 map of CS, spatially matching the observed streamer emission to the north (outlined in the green polygon). Bottom: Velocity components of CS taken from inside the green polygon plotted as a function of projected distance from the central young stellar object, with varying levels of kernel density estimation of the velocity plotted as filled contours.  The contours start at 0.5$\sigma$ and progress in steps of 0.5$\sigma$, where $\sigma$ is from a bivariate normal distribution \citep[see also][]{Pineda2020}. The streamline model, shown in green, describes most of the kinematics of the system in addition to the spatial match shown in the top panel. The main deviation is a small amount of blue-shifted emission in the streamer at a projected distance of $\sim$100-150 au from the central young stellar object.}
 \label{DGTau_streamer}
 \end{figure}

{DG Tau has a northern streamer traced in CS and well-modeled both spatially and kinematically by an infalling streamline model (see Fig.~\ref{DGTau_streamer}). The infalling streamer lands in the disk where there is SO emission (compare to Fig.~\ref{Imagery_DGTau}). The correspondence between the streamer impact zone and the SO emission indicates that the SO is likely tracing an accretion shock of material entering the disk from the envelope through the streamer.}

\begin{figure}
  \centering
 \includegraphics[width=9cm]{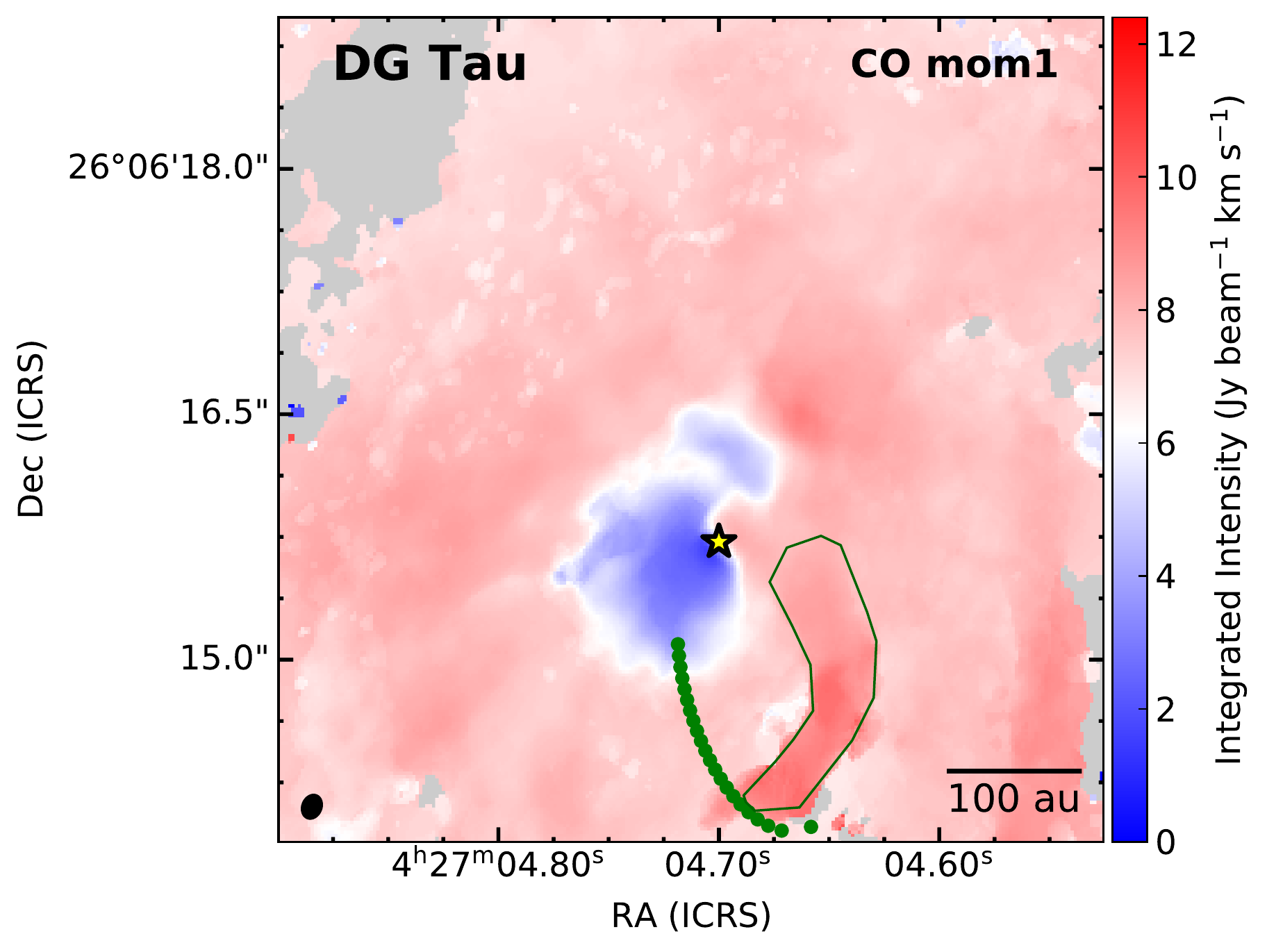}
 \includegraphics[width=9cm]{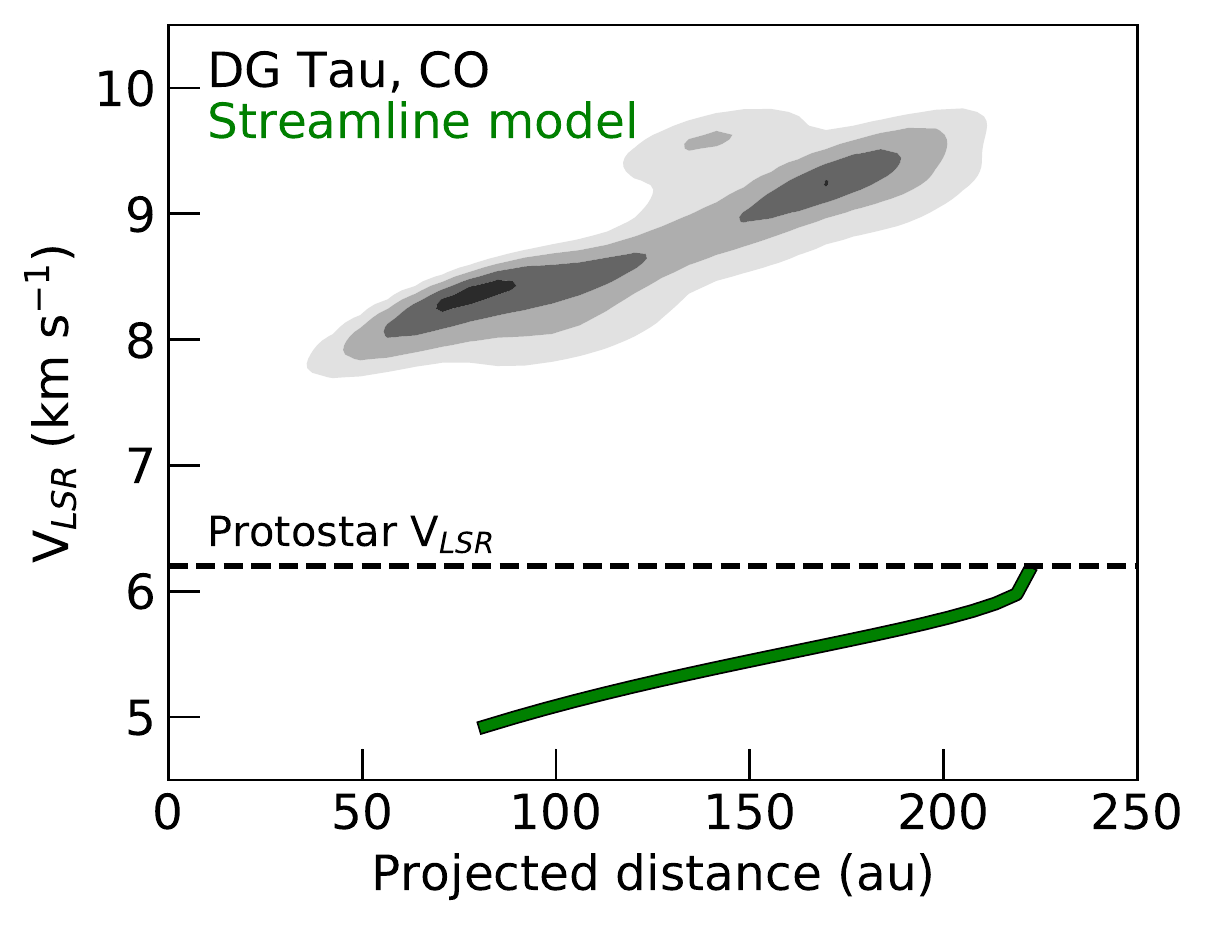}
     \caption{{Infalling streamline model for the red-shifted southern arc structure (outlined in the green polygon), with the same disk orientation and rotation geometry used for the northern streamer model (Fig.~\ref{DGTau_streamer}). Top: Streamline model, plotted in green, on top of the moment-1 map of CO, showing the model curvature with this rotation geometry is opposite to the spatial structure of the southern redshifted arc. Bottom: Velocity components of CO taken from inside the green polygon plotted as a function of projected distance from the central young stellar object, with varying levels of kernel density estimation of the velocity plotted as filled contours and contours calculated as in Fig.~\ref{DGTau_streamer}. The streamline model consistently under-predicts the velocity of the arc structure by $\sim$3 km s$^{-1}$.}}
 \label{DGTau_streamer-south}
 \end{figure}

\begin{figure}
  \centering
 \includegraphics[width=9cm]{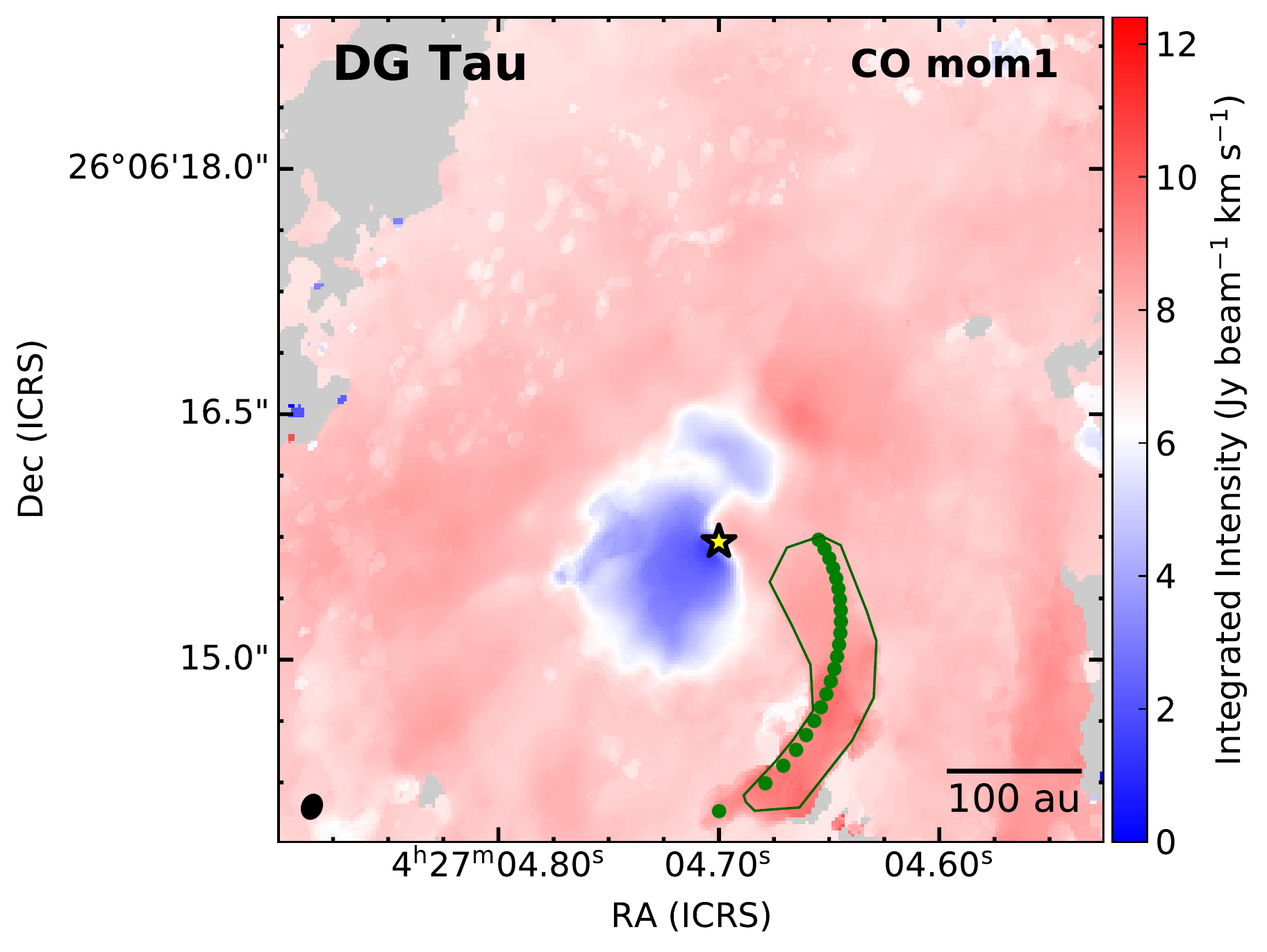}
 \includegraphics[width=9cm]{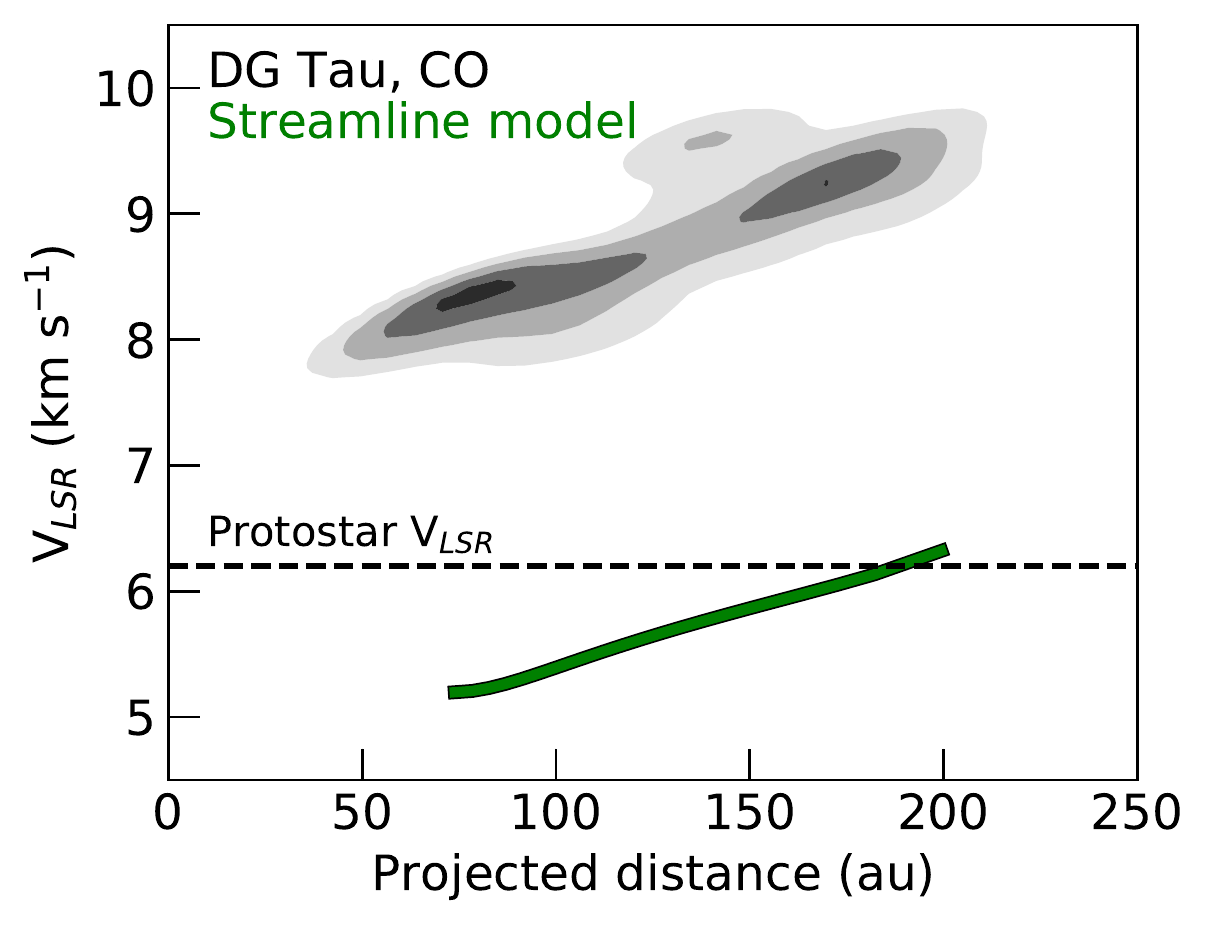}
     \caption{{Infalling streamline model for the redshifted southern arc structure (outlined in the green polygon), with the same disk orientation but opposite rotation geometry used for Fig.~\ref{DGTau_streamer-south}. Top: Streamline model, plotted in green, on top of the moment-1 map of CO, showing the model curvature well matches spatial structure of the southern redshifted arc with the envelope rotation counter-rotating with respect to the disk. Bottom: Velocity components of CO taken from inside the green polygon plotted as a function of projected distance from the central young stellar object, with varying levels of kernel density estimation of the velocity plotted as filled contours and contours calculated as in Fig.~\ref{DGTau_streamer}. Here, the streamline model also consistently under-predicts the velocity of the arc structure by $\sim$3 km s$^{-1}$, even though the spatial structure is well matched by the model.}}
 \label{DGTau_streamer-south-180}
 \end{figure}

{{Conversely, the redshifted southern arc is not well described by the streamline model. In Figures \ref{DGTau_streamer-south} and \ref{DGTau_streamer-south-180}, we show our two best attempts to model the redshifted southern arc in CO with different rotational geometries.  For both models, we leave the disk orientation the same as it was in the modeling for the northern streamer traced in CS.  For Figure B.2, we use a rotational geometry where the envelope co-rotates with the disk, which we have used for the other successful streamline models in this work and which would be expected from a simple scenario where the disk formed within a rotating envelope.  The model parameters $M_{star}$, $V_{lsr}$ $I.A._{disk}$, and $P.A._{disk}$ remain the same as reported in Table \ref{tab:stream_params}, and $r_0$ = 300 au, $\theta_0$ = 110$^{\circ}$, $\phi_0$ = 60$^{\circ}$, $\Omega_0$ = $1.0 \times 10^{-11}$ s$^{-1}$ and $v_{r,0}$ = 0.0 km s$^{-1}$.  The streamline model curvature is oppositely curved with respect to the structure of the redshifted arc, and the velocities are under-predicted by $\sim$3 km s$^{-1}$.  For Figure \ref{DGTau_streamer-south-180}, we show a model where the envelope counter-rotates with respect to the disk, which could be the case if the redshifted southern arc were an infalling captured cloudlet.   Here, $I.A._{disk}$ and $P.A._{disk}$ have rotations applied to maintain the same physical orientation of the disk while changing the rotation direction of the infalling streamers to counter-rotate with respect to the disk. $r_0$ = 300 au, $\theta_0$ = 80$^{\circ}$, $\phi_0$ = 110$^{\circ}$, $\Omega_0$ = $6.0 \times 10^{-12}$ s$^{-1}$ and $v_{r,0}$ = 0.0 km s$^{-1}$. In this case, while the spatial curvature of the arc is well matched, the velocities remain under-predicted by $\sim$3 km s$^{-1}$.  }}

{Considering that the redshifted southern arc seems to have a coherent velocity structure with the northern streamer, this may indicate that material from the northern streamer may have enough momentum to pass through the disk midplane and is beginning to curve back towards the young stellar object in the redshifted arc. Hence, the redshfited arc may be a putative southern extension of the northern streamer.  The analytic streamline model would be unable to model such a complex scenario where material passes through the midplane, because the model assumes that any material reaching the midplane will be met by material in a mirror-opposite streamline from the opposing hemisphere and cannot therefore  vertically pass through the disk \citep{Mendoza2009}. The redshifted arc could also be a cloudlet in the process of being captured that is moving with a trajectory not well described by the analytic streamline model {or potentially disturbed by the outflow}.}

\subsection{HL Tau}\label{App_HLTau}
\begin{figure}
  \centering
 \includegraphics[width=9cm]{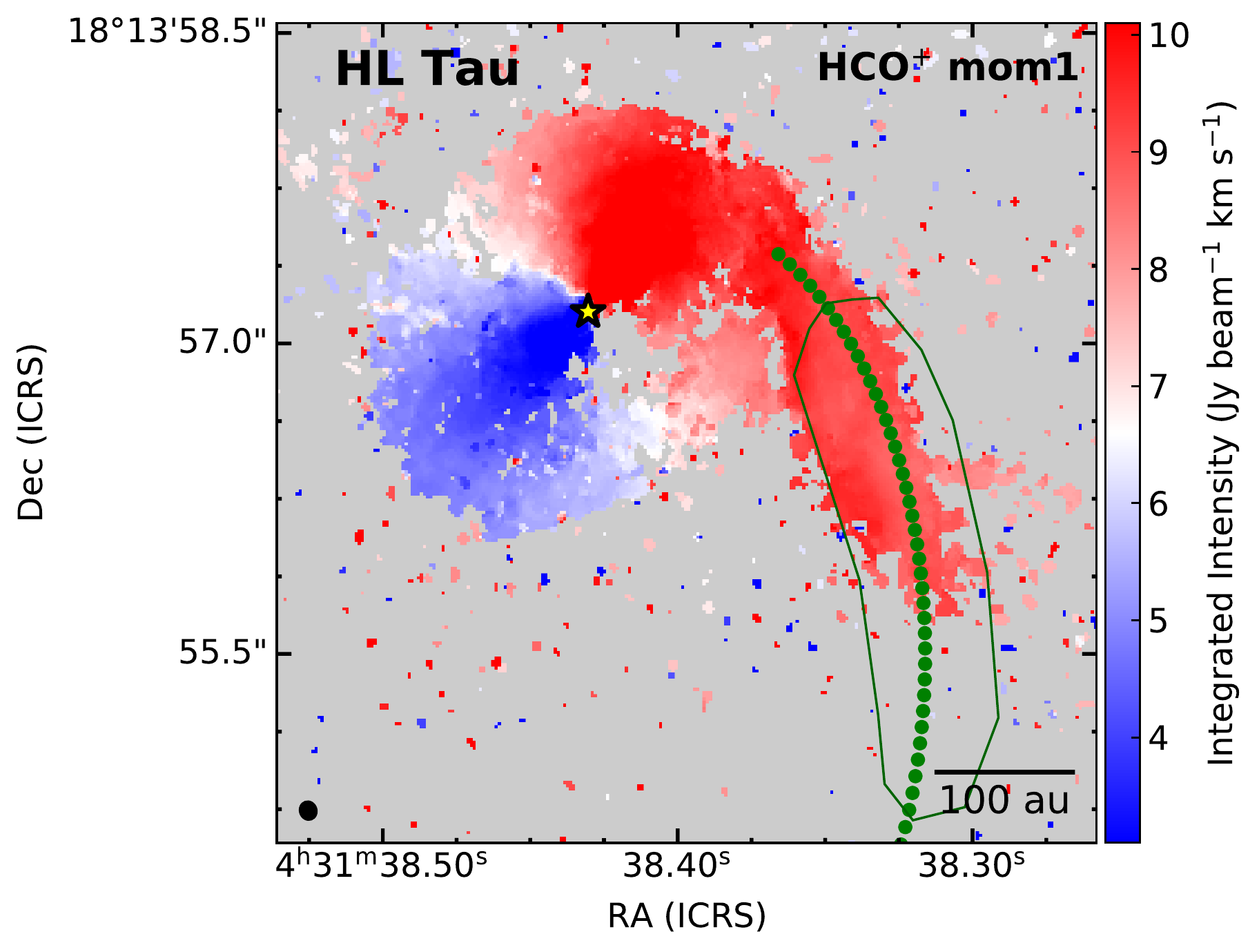}
 \includegraphics[width=9cm]{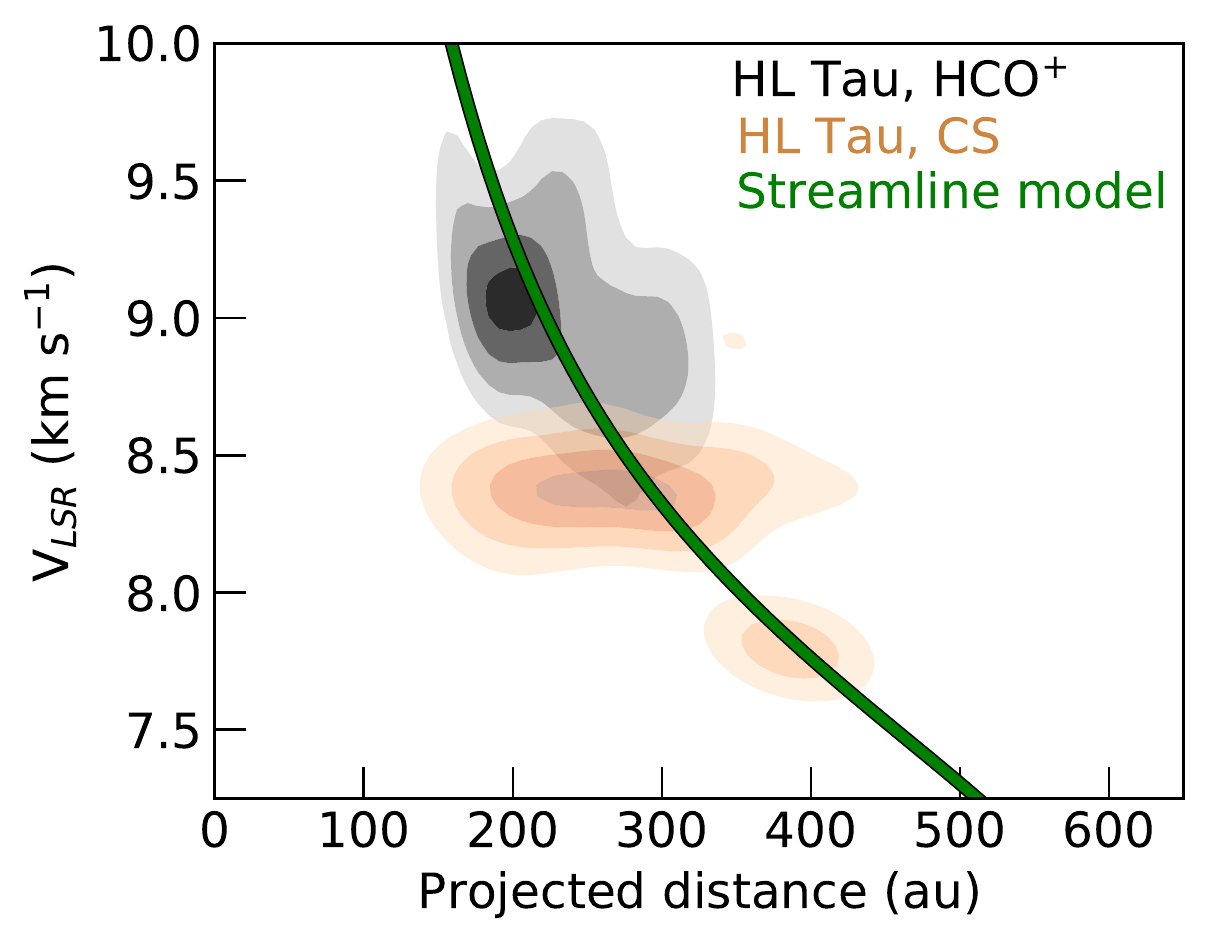}
     \caption{Infalling streamline model for the HL Tau streamer. Top: Streamline model, plotted in green, on top of the moment-1 map of HCO$^{+}$, spatially matching the observed streamer emission extending to the south (outlined in the green polygon). The model predicts that the streamer should extend beyond the HCO$^{+}$ emission, into a region traced by CS (see Fig.~\ref{Imagery_HLTau}). Bottom: Velocity components of HCO$^{+}$ (black) and CS (orange) taken from inside the green polygon plotted as a function of projected distance from the central young stellar object, with varying levels of kernel density estimation of the velocity plotted as filled contours and contours calculated as in Fig.~\ref{DGTau_streamer}. The streamline model is shown in green, and highlights that both HCO$^{+}$ and CS trace different parts of the same infalling streamer.}
 \label{HLTau_streamer}
 \end{figure}

{The streamer of HL Tau to the southwest of the disk is traced by HCO$^+$ near the disk, with radii further from the disk traced by CS. Both the HCO$^+$ and CS in the streamer are well described by the analytic infalling streamline model (see Fig.~\ref{HLTau_streamer}). This is the first time a single streamer has been traced and modeled with different molecular tracers at different projected distances from the system center. The streamer lands where the SO extends to the west as seen in Fig.~\ref{Imagery_HLTau}, tracing the shocked impact zone of the streamer.}

\subsection{IRAS 04302+2247}\label{App_IRAS}
\begin{figure}
  \centering
 \includegraphics[width=9cm]{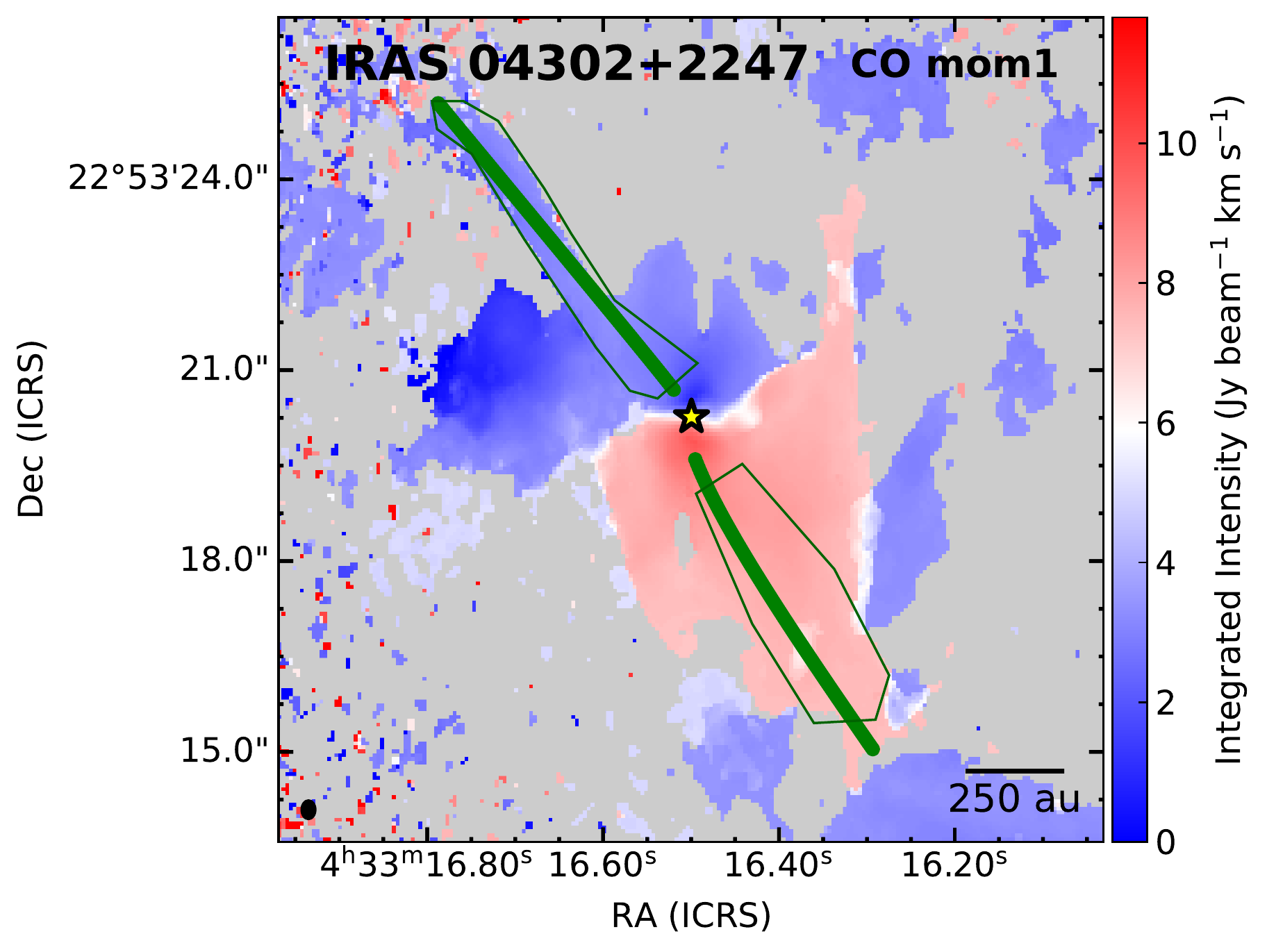}
 \includegraphics[width=9cm]{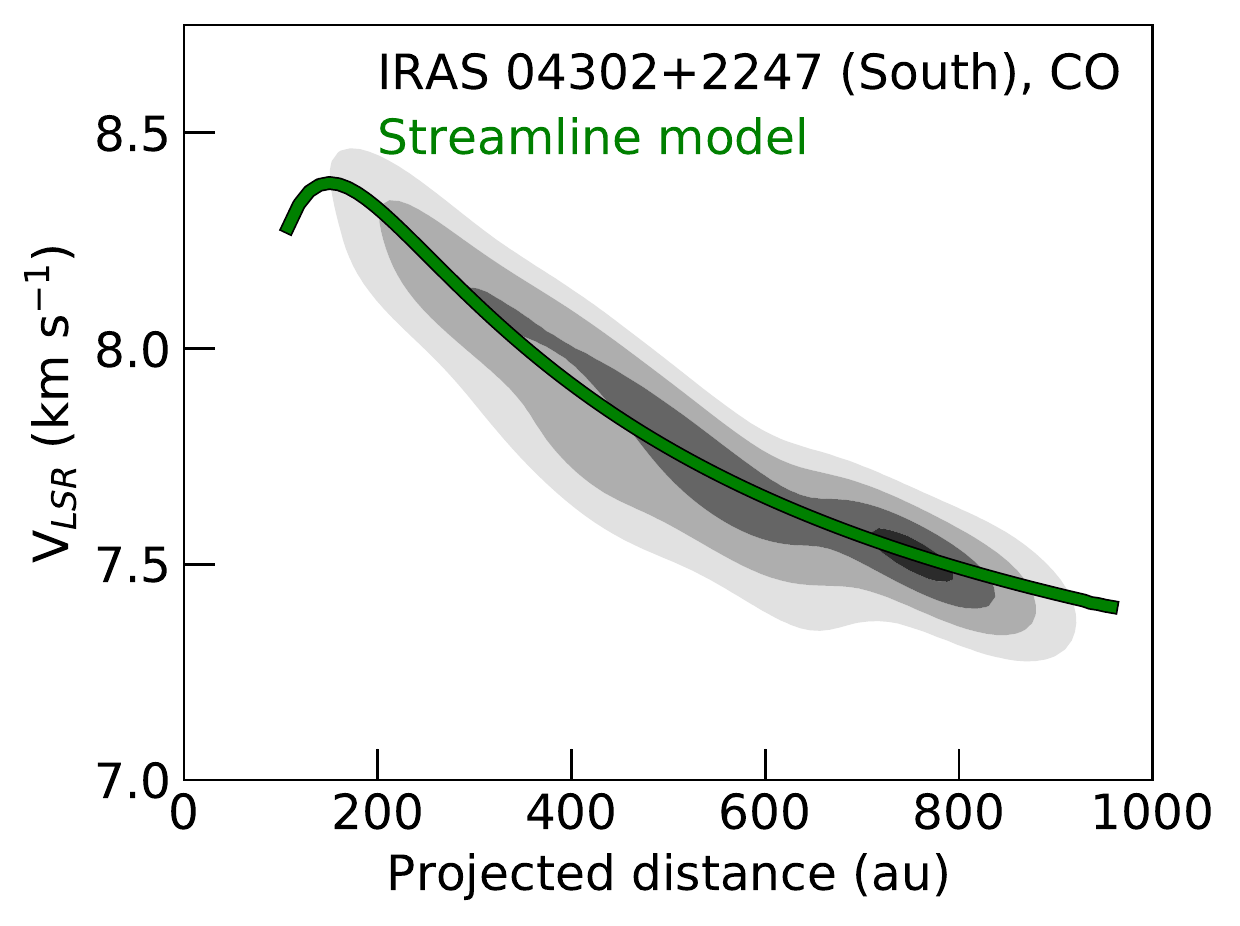}
 \includegraphics[width=9cm]{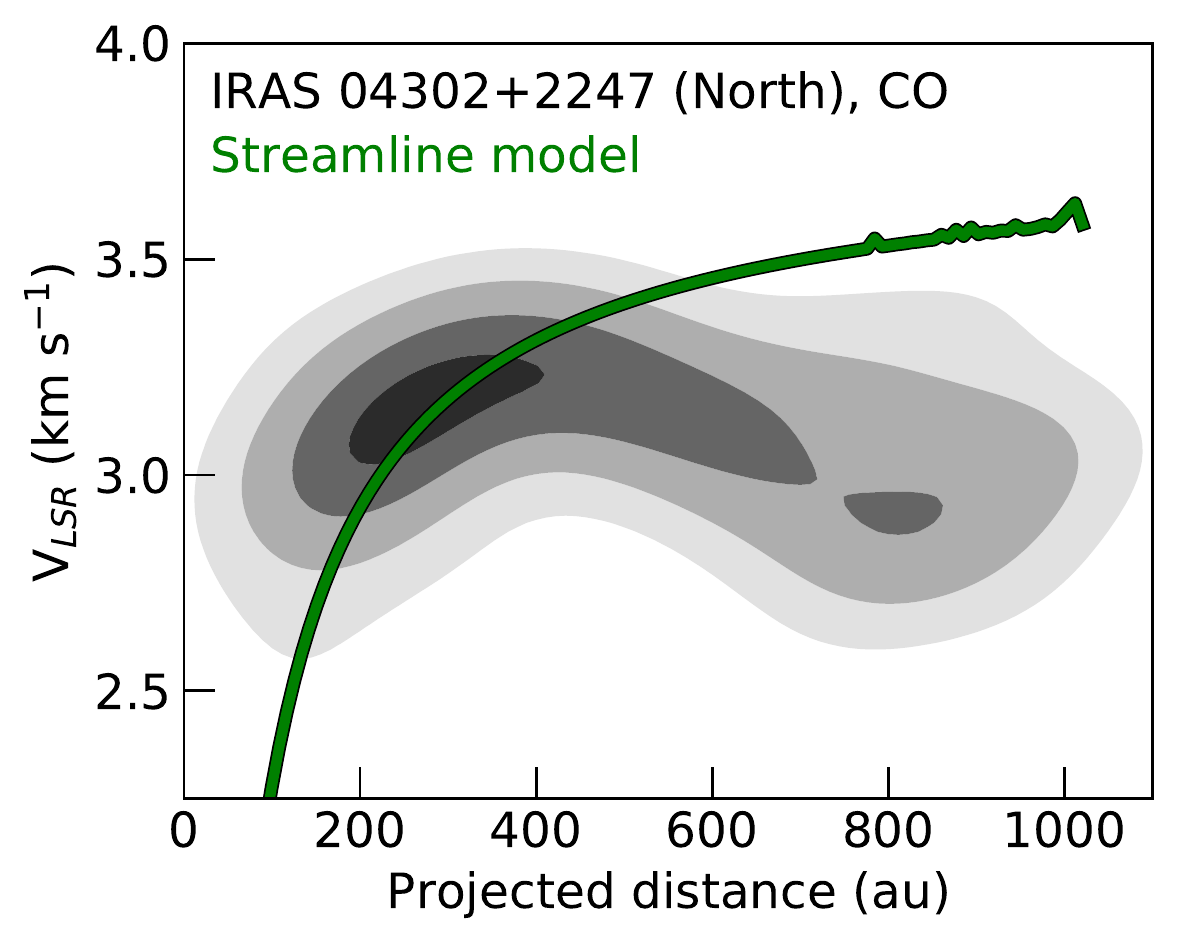}
     \caption{Infalling streamline model for the two IRAS04302 streamers. Top: Streamline models, plotted in green, on top of the moment-1 map of CO, spatially matching the observed streamer emission to both the south and north (outlined in the green polygons). Middle and bottom: Velocity components of CO taken from inside the south and north green polygons, respectively, plotted as a function of projected distance from the central young stellar object, with varying levels of kernel density estimation of the velocity plotted as filled contours and contours calculated as in Fig.~\ref{DGTau_streamer}. The streamline models are shown in green. For the southern streamer, the CO emission is entirely described by the streamline model.  For the northern streamer, the streamline model corresponds well with the emission at radii less than 400 au, but the model over-predicts the velocity at larger radii, hinting at a more complex velocity field in this system.}
 \label{IRAS_streamer}
 \end{figure}

{IRAS 04302 has two infalling streamers that can be identified by the analytic streamline modeling of the CO emission (Fig.~\ref{IRAS_streamer}).  The redshifted southern streamer seems to feed from the blueshifted captured cloudlet, with SO present in the disk where the model predicts the impact zone of the streamer to be (compare Fig.~\ref{IRAS_streamer} and Fig.~\ref{Imagery_IRAS}). The blueshifted northern streamer is well described by the streamline model at small radii, though at larger radii the model over-predicts the velocity of the observed material, and suggests that the larger scale environment around the disk is likely much more complicated than the simple streamline model can account for.}

\end{appendix}

\end{document}